\documentclass[aps,twocolumn,prd,showpacs,showkeys,preprintnumbers,superscriptaddress,nobibnotes,floatfix,longbibliography,nofootinbib]{revtex4-2}
\pdfoutput=1
\usepackage{amsmath}
\usepackage{amsfonts}
\usepackage{amssymb}
\usepackage{mathrsfs}
\usepackage{graphicx}
\usepackage{color}
\usepackage{hyperref}
\usepackage{multirow}
\usepackage{slashed}
\usepackage{epsf}

\usepackage{graphicx}
\usepackage{caption}
\usepackage{subcaption}

\usepackage{color}

\hypersetup{
   colorlinks=true,       
   linkcolor=blue,        
   citecolor=red,         
   filecolor=magenta,      
   }

\def\to{\rightarrow}

\newcommand{\Rmnum}[1]{\MakeUppercase{\romannumeral #1}}

\begin{document}

\title{
Revisiting the Electroweak Supersymmetry from the Generalized Minimal Supergravity
}

\author{Imtiaz Khan}
\email{ikhanphys1993@gmail.com}
\affiliation{Department of Physics, Zhejiang Normal University, Jinhua, Zhejiang 321004, China}
\affiliation{Zhejiang Institute of Photoelectronics, Jinhua, Zhejiang 321004, China}

\author{Ali Muhammad}
\email{alimuhammad@phys.qau.edu.pk}
\affiliation{CAS Key Laboratory of Theoretical Physics, Institute of Theoretical Physics, Chinese Academy of Sciences, Beijing 100190, China}
\affiliation{School of Physical Sciences, University of Chinese Academy of Sciences, No. 19A Yuquan Road, Beijing 100049, China}

\author{Tianjun Li}
\email{tli@mail.itp.ac.cn}
\affiliation{School of Physics, Henan Normal University, Xinxiang 453007, P. R. China}
\affiliation{CAS Key Laboratory of Theoretical Physics, Institute of Theoretical Physics, Chinese Academy of Sciences, Beijing 100190, China}
\affiliation{School of Physical Sciences, University of Chinese Academy of Sciences, No. 19A Yuquan Road, Beijing 100049, China}

\author{Shabbar Raza}
\email{shabbar.raza@fuuast.edu.pk}
\affiliation{Department of Physics, Federal Urdu University of Arts, Science and Technology, Karachi 75300, Pakistan}

\begin{abstract}

We explore the Electroweak Supersymmetry (EWSUSY) scenario within the Minimal Supersymmetric Standard Model (MSSM) under the Generalized Minimal Supergravity (GmSUGRA) framework, given that the anomalous magnetic moment of the muon may now be consistent with the Standard Model (SM) prediction, we consider both signs of the Higgsino mass parameter, \( \mu < 0 \) and \( \mu > 0 \). A comprehensive scan of the parameter space is performed, subject to the experimental constraints from  the LHC SUSY searches, Planck 2018 relic density, and LUX-ZEPLIN (LZ) direct detection limits. We identify the viable regions featuring neutralino dark matter production via coannihilation with stau, chargino, stop, sbottom, and gluino, as well as through $A$-funnel, Higgs-resonance, and $Z$-resonance mechanisms. Notably, the $\mu < 0$ scenario yields a broader allowed parameter space, including for the first time sbottom-neutralino coannihilation solutions in GmSUGRA, which are absent for $\mu > 0$. While the Higgs-pole and $Z$-pole regions for $\mu > 0$ are largely excluded by the current LZ bounds, substantial viable regions remain for $\mu < 0$. Gluino coannihilation scenarios are strongly constrained by the current LHC data. The characteristic mass ranges of interest include sbottoms (0.7--1.3~TeV), stops (up to 1.0~TeV for $\mu > 0$ and 1.3~TeV for $\mu < 0$), staus and charginos (up to 1.5~TeV), and pseudoscalar Higgs bosons in the $A$-funnel (0.4--1.4~TeV). Moreover, the supersymmetric contributions to the muon anomalous magnetic moment remain within a \(2\sigma\) deviation from the SM prediction. And
 our findings suggest that significant portions of the parameter space can be probed at the future LHC SUSY searches and upcoming dark matter direct detection experiments.
 
\end{abstract}

\maketitle


\textbf{Introduction:}
\label{intro}
The Supersymmetric Standard Models (SSMs) remain among the most attractive frameworks for physics beyond the Standard Model (BSM). They offer elegant solutions to several theoretical challenges, such as the gauge hierarchy problem , and predict gauge coupling unification~\cite{gaugeunification,Georgi:1974sy,Pati:1974yy,Mohapatra:1974hk,Fritzsch:1974nn,Georgi:1974my}. Moreover, if $R$-parity is conserved, the lightest supersymmetric particle (LSP), often the neutralino, constitutes a viable dark matter (DM) candidate~\cite{neutralinodarkmatter,darkmatterreviews}. The framework also naturally accommodates radiative electroweak symmetry breaking via the large top Yukawa coupling. Within the Minimal Supersymmetric Standard Model (MSSM), the predicted mass for the lightest CP-even Higgs boson lies in the 100--135~GeV range~\cite{Slavich:2020zjv}, in agreement with experimental observations. As a result, supersymmetry (SUSY) remains a central focus at the Large Hadron Collider (LHC), linking high-scale theory with low-energy phenomenology.
Following the LHC Run-2, no definitive SUSY signals have emerged, leading to stringent mass bounds on colored sparticles: $\sim$1.25~TeV for stops, $\sim$1.5~TeV for sbottoms, $\sim$2~TeV for first- and second-generation squarks, and $\sim$2.2~TeV for gluinos~\cite{ATLAS-SUSY-Search,Aad:2020sgw,Aad:2019pfy,CMS-SUSY-Search-I,CMS-SUSY-Search-II}. These results suggest that colored superpartners are considerably heavy, typically above the TeV scale. However, viable models with an electroweak-scale bino-like LSP and relatively light sleptons remain consistent with current constraints. Motivated by BSM theories, such scenarios allow slepton masses ranging from a few hundred GeV up to the TeV  scale~\cite{Ahmed:2022ude,Zhang:2023jcf,Khan:2023ryc}.

Recent advances in collider and dark matter (DM) experiments have significantly refined our understanding of physics BSM. The ATLAS and CMS collaborations have provided important constraints through searches for heavy Higgs bosons~\cite{ATLAS:2020zms}, detailed studies of charginos and neutralinos~\cite{CMS:2020bfa,ATLAS:2021moa,ATLAS:2021yqv,CMS:2022sfi}, and investigations of invisible Higgs decays~\cite{ATLAS:2022yvh}. In parallel, direct detection (DD) experiments such as PICO-60, LUX-ZEPLIN (LZ), XENON-1T, and PandaX-4T have set stringent limits on spin-independent (SI) and spin-dependent (SD) DM-nucleon scattering cross-sections~\cite{XENON:2018voc,XENON:2019rxp,PICO:2019vsc,PandaX-4T:2021bab,LZ:2022lsv,LZ:2024zvo,PandaX:2022xas}. With the Large Hadron Collider (LHC) entering Run-3, a timely reassessment of the MSSM parameter space is essential. In light of the most recent and stringent SI cross-section bounds reported by the LZ experiment~\cite{LZ:2022lsv,LZ:2024zvo}, identifying phenomenologically viable regions that remain accessible to upcoming LHC searches is of critical importance.

We present a detailed phenomenological study of the Generalized Minimal Supergravity (GmSUGRA) framework, investigating both signs of the higgsino mass parameter, $\mu > 0$ and $\mu < 0$, without relying on any fine-tuning assumptions. The sign of $\mu$ plays a pivotal role in supersymmetric (SUSY) Grand Unified Theories (GUTs), particularly in relation to the longstanding deviation in the anomalous magnetic moment of the muon, $(g_\mu - 2)$. Assuming SUSY manifests at the electroweak scale, the majority of  SUSY 
contribution to the muon anomalous magnetic moment, $a_\mu$, primarily appears from loop diagrams involving neutralino-smuon  and chargino-sneutrino interactions. These contributions can be approximately described by~\cite{KhalilS2017}:
\begin{equation}
\Delta a_\mu^{\text{SUSY}} \sim \frac{M_i \, \mu \, \tan\beta}{m_{\text{SUSY}}^4},
\label{eq:g-2}
\end{equation}
where $M_i$ ($i=1,2$) are the electroweak gaugino masses, $\mu$ denotes the Higgsino masses parameter, $\tan\beta = \langle H_u \rangle / \langle H_d \rangle$ is the ratio of Higgs vacuum expectation values, and $m_{\text{SUSY}}$ represents the characteristic mass scale of SUSY particles running in the loop. It is evident from Eq.~\ref{eq:g-2} that a positive $\mu$ enhances the SUSY contribution to $a_\mu$, especially in scenarios with light sleptons and electroweakinos, and is therefore favored by current $(g_\mu - 2)$ data~\cite{Muong-2:2023cdq, Ahmed:2021htr}. Conversely, in scenarios where the $(g_\mu - 2)$ anomaly is not considered or becomes less significant, the $\mu < 0$ case gains renewed importance. A new and precise determination of the anomalous magnetic moment of the positive muon, \( a_\mu \), has been reported by the Muon \((g{-}2)\) Experiment at Fermilab (E989), based on data collected between 2020 and 2023 \cite{Muong-2:2023cdq,Muong-2:2021ojo}. From this latest dataset, the experiment finds \cite{Muong-2:2025xyk}:
\[
a_\mu = 116\,592\,0710(162) \times 10^{-12} \quad \text{(139 ppb)}.
\]
Combining this result with previous measurements yields a refined value\cite{Muong-2:2025xyk}:
\[
a_\mu = 116\,592\,0705(148) \times 10^{-12} \quad \text{(127 ppb)}.
\]
The updated world average, which is now predominantly driven by Fermilab data, is given by\cite{Muong-2:2025xyk}:
\[
a_\mu^{\text{exp}} = 116\,592\,0715(145) \times 10^{-12} \quad \text{(124 ppb)}.
\]
This represents a factor of four improvement in the experimental precision compared to earlier results.
Simultaneously, significant advancements have been achieved in lattice QCD computations of the leading-order hadronic vacuum polarization (LO HVP) contribution. The consolidated lattice-QCD estimate now attains a precision of approximately 0.9\%, enabling a substantially more accurate Standard Model (SM) prediction \cite{Aliberti:2025beg}. Incorporating this refined input, the current SM prediction for the muon anomaly is \cite{Aliberti:2025beg}:
\[
a_\mu^{\text{SM}} = 116\,592\,033(62) \times 10^{-11} \quad \text{(530 ppb)}.
\]
The difference between the experimental average and the updated SM prediction is now \cite{Muong-2:2025xyk,Aliberti:2025beg}:
\[
\Delta a_\mu = a_\mu^{\text{exp}} - a_\mu^{\text{SM}} = 38.5(63.673) \times 10^{-11},
\]
which corresponds to a statistical deviation of merely \( 0.6\sigma \). This result indicates that the previously reported tension between experiment and theory has effectively been resolved at the current level of experimental and theoretical precision.
Moreover, recent studies highlight that it $\mu < 0$ provides viable DM solutions via the Higgs- and $Z$-boson resonance (pole) channels while simultaneously remaining consistent with direct detection and LHC constraints~\cite{Barman:2022jdg, Khan:2025azf}. All these finding push us to the phenomenological studies of SUSY GUTs in light of the current LHC SUSY searches and DD DM experiments for the $\mu<0$ scenario.

Focusing on dark matter phenomenology, we conduct extensive scans over the GmSUGRA parameter space and identify regions compatible with current limits from direct detection experiments such as the LUX-ZEPLIN (LZ) Collaboration. Both $\mu > 0$, and $\mu < 0$ scenarios yield solutions featuring resonant annihilation via the Higgs and $Z$ bosons. However, direct detection bounds impose significantly tighter constraints on the $\mu > 0$ case, largely excluding light higgsinos in the $Z$- and $H$-pole regions. In contrast, the $\mu < 0$ case allows for such light higgsinos to remain viable under current LZ and electroweakino search limits. Our analysis uncovers a wide range of DM mechanisms compatible with collider and cosmological data, including: $A$-resonance annihilation, Higgs- and $Z$-boson resonance channels, as well as coannihilation scenarios involving stop, stau, gluino, sbottom, and tau with the lightest neutralino. All identified solutions respect constraints from LHC Run 2 data, projected sensitivities of Run 3, and future high-energy collider prospects. Moreover, these scenarios are consistent with the relic density bounds reported by Planck 2018 and adhere to current and forthcoming limits from direct and indirect DM detection efforts. Significantly, for the first time within the GmSUGRA framework, we identify sbottom-neutralino coannihilation as a viable mechanism exclusively in the $\mu < 0$ scenario, whereas this channel is absent for $\mu > 0$. This further supports the case for investigating $\mu < 0$ more thoroughly, given its greater flexibility under existing experimental bounds. Our results also demonstrate that the supersymmetric contribution to the muon's anomalous magnetic moment, \( (g{-}2)_\mu \), remains compatible with the current experimental measurement, yielding deviations within the \(1\sigma\) to \(2\sigma\) deviation around the SM expectation for the $\mu < 0$ scenario. Thus, our findings underscore the phenomenological richness of the GmSUGRA model and motivate focused studies on the $\mu < 0$ regime to draw more conclusive insights in light of ongoing and future experimental results.

\section{The GmSUGRA in the MSSM}
\label{model}
Electroweak supersymmetry (EWSUSY) can naturally emerge within the framework of the GmSUGRA model~\cite{Li:2010xr, Balazs:2010ha}, where the sleptons, bino, wino, and/or higgsinos lie within the TeV scale, while squarks and gluinos may reside at multi-TeV scales~\cite{Cheng:2012np}. The GmSUGRA is based on the $SU(5)$ GUT, with the GUT symmetry broken by a Higgs field in the adjoint representation. Couplings of this field to gauge kinetic terms via higher-dimensional operators induce modifications in the gauge coupling and gaugino mass unification conditions once the adjoint acquires a vacuum expectation value (VEV).

The modified gauge coupling and gaugino mass relations at the GUT scale are given by:
\begin{align}
\frac{1}{\alpha_2} - \frac{1}{\alpha_3} &= k\left( \frac{1}{\alpha_1} - \frac{1}{\alpha_3} \right), \\
\frac{M_2}{\alpha_2} - \frac{M_3}{\alpha_3} &= k\left( \frac{M_1}{\alpha_1} - \frac{M_3}{\alpha_3} \right),
\end{align}
where \( k = 5/3 \) in the minimal GmSUGRA case. Assuming gauge coupling unification, \( \alpha_1 = \alpha_2 = \alpha_3 \), leads to a simplified gaugino mass relation:
\begin{equation}
M_2 - M_3 = \frac{5}{3}(M_1 - M_3),
\label{M3a}
\end{equation}
which generalizes the universal gaugino mass condition \( M_1 = M_2 = M_3 \) of minimal supergravity (mSUGRA). Solving Eq.~\eqref{M3a} for \( M_3 \), we obtain:
\begin{equation}
M_3 = \frac{5}{2}M_1 - \frac{3}{2}M_2,
\label{M3}
\end{equation}
implying that \( M_3 \) can vary from several hundred GeV to multiple TeV depending on the values of \( M_1 \) and \( M_2 \), which are treated as free input parameters in EWSUSY. The soft supersymmetry-breaking (SSB) scalar mass relations at the GUT scale are derived under the assumption of an adjoint Higgs in $SU(5)$~\cite{Balazs:2010ha}. Taking the slepton masses as free parameters, the squark masses are:
\begin{align}
m_{\tilde{Q}_i}^2 &= \frac{5}{6} (m_0^U)^2 + \frac{1}{6} m_{\tilde{E}_i^c}^2, \\
m_{\tilde{U}_i^c}^2 &= \frac{5}{3} (m_0^U)^2 - \frac{2}{3} m_{\tilde{E}_i^c}^2, \\
m_{\tilde{D}_i^c}^2 &= \frac{5}{3} (m_0^U)^2 - \frac{2}{3} m_{\tilde{L}_i}^2,
\label{squarks_masses}
\end{align}
where \( m_0^U \) is the universal scalar mass, and the tilde fields represent soft scalar masses of the corresponding sfermions. In EWSUSY, since \( m_{\tilde{L}}, m_{\tilde{E}^c} \lesssim 1~\mathrm{TeV} \), the sleptons remain light. In the limit \( m_0^U \gg m_{\tilde{L}}, m_{\tilde{E}^c} \), the approximate relations \( 2m_{\tilde{Q}}^2 \approx m_{\tilde{U}^c}^2 \approx m_{\tilde{D}^c}^2 \) hold....
Additionally, the Higgs soft masses \( m_{\tilde{H}_u} \), \( m_{\tilde{H}_d} \), and trilinear couplings \( A_U, A_D, A_E \) are all considered free parameters in the GmSUGRA framework~\cite{Balazs:2010ha, Cheng:2012np}.

\section{Phenomenological Constraints and Scanning Procedure}
\label{sec:scan}
We perform a random scan over the parameter space using the \texttt{ISAJET~7.85} package~\cite{ISAJET}. The scan is carried out over the following ranges:
	\begin{align}
	0 \, \rm{GeV} \leq & m_0^{U}  \leq 9000 \, \rm{GeV}  ~,~\nonumber \\
	80 \, \rm{GeV} \leq & M_1  \leq 3000 \, \rm{GeV} ~,~\nonumber \\
	100\, \rm{GeV} \leq & M_2   \leq 3100 \, \rm{GeV} ~,~\nonumber \\
	100 \, \rm{GeV} \leq & m_{\tilde L}  \leq 1200 \, \rm{GeV} ~,~\nonumber \\
	100 \, \rm{GeV} \leq & m_{\tilde E^c}  \leq 1200 \, \rm{GeV} ~,~\nonumber \\
	100 \, \rm{GeV} \leq & m_{\tilde H_{u,d}} \leq 5000 \, \rm{GeV} ~,~\nonumber \\
	-16000 \, \rm{GeV} \leq & A_{U}=A_{D} \leq 16000 \, \rm{GeV} ~,~\nonumber \\
	-6000 \, \rm{GeV} \leq & A_{E} \leq 6000 \, \rm{GeV} ~,~\nonumber \\
	2\leq & \tan\beta  \leq 60~.~
\label{eqn:scan}
	\end{align}

The scan assumes \( \mu > 0 \), \( \mu < 0 \), and a top quark mass of \( m_t = 173.3\, \text{GeV} \)~\cite{ATLAS:2014wva}, with \( m_b^{\overline{\text{DR}}}(M_Z) = 2.83\, \text{GeV} \) as implemented in \texttt{ISAJET}. Parameter space exploration is guided by the Metropolis-Hastings algorithm~\cite{Belanger:2009ti}. Throughout the analysis, we adopt the notations \( A_t \equiv A_U \), \( A_b \equiv A_D \), and \( A_\tau \equiv A_E \).

The collected data is subjected to a set of experimental and theoretical constraints discussed in the subsequent section.

\noindent
\textbf{(\Rmnum{1}) Basic Constraints:}
\begin{itemize}
    \item[(a)] Radiative Electroweak Symmetry Breaking (REWSB).
    \item[(b)] The lightest supersymmetric particle (LSP) is a neutralino.
    \item[(c)] Sparticle mass bounds:
    \begin{align}
    m_{\tilde{t}_1},\ m_{\tilde{b}_1} &\geq 100\, \text{GeV}, \nonumber \\
    m_{\tilde{\tau}_1} &\geq 105\, \text{GeV}, \nonumber \\
    m_{\tilde{\chi}_1^{\pm}} &\geq 103\, \text{GeV},
    \label{eqn:spMassLEP2}
    \end{align}
    from LEP2 \cite{Patrignani:2016xqp}, and from LHC:
    \begin{align}
    m_{\tilde{g}} &\geq 2.2\, \text{TeV}, \nonumber \\
    m_{\tilde{q}} &\geq 2\, \text{TeV}, \nonumber \\
    m_{\widetilde t_1}& \gtrsim ~ 1.25 \text{TeV},\nonumber \\
    m_{\widetilde b_1} &\gtrsim ~ 1.5 \,\text{TeV}
    \label{eqn:spMassLHC}
    \end{align}
    with references~\cite{ATLAS-SUSY-Search,Aad:2020sgw,Aad:2019pfy,CMS-SUSY-Search-I,CMS-SUSY-Search-II}.
    \item[(d)] Higgs boson mass:
    \begin{equation}
    122\, \text{GeV} \leq m_h \leq 128\, \text{GeV}, \label{eqn:higgsMassLHC}
    \end{equation}
     to account for the uncertainty of $m_h$ calculation in the MSSM with refrences~\cite{Khachatryan:2016vau,Allanach:2004rh}.
 \item[(e)] \textit{B}-physics observables (evaluated using \texttt{IsaTools}~\cite{Baer:1997jq,Babu:1999hn,Dedes:2001fv,Mizukoshi:2002gs}):
    \begin{align}
    & 0.15 \leq \frac{
 {\rm BR}(B_u\rightarrow\tau \nu_{\tau})_{\rm MSSM}}
 {{\rm BR}(B_u\rightarrow\tau \nu_{\tau})_{\rm SM}}
        \leq 2.41 \; (3\sigma)~,
\\
& 2.99 \times 10^{-4} \leq
  {\rm BR}(b \rightarrow s \gamma)
  \leq 3.87 \times 10^{-4} \; (2\sigma)~,
\\
& 0.8\times 1-^{-9} \leq{\rm BR}(B_s \rightarrow \mu^+ \mu^-)
  \leq 6.2 \times10^{-9} \;(2\sigma). \label{eqn:Butaunu}
    \end{align}
    within 3$\sigma$ and 2$\sigma$ experimental limits~\cite{Aaij:2012nna,Amhis:2012bh,Asner:2010qj}.
    \item[(f)] Planck 2018 relic density constraint~\cite{Akrami:2018vks}:
    \begin{equation}
    0.114 \leq \Omega_{\rm CDM}h^2 (\rm Planck2018) \leq 0.126   \; (5\sigma)
     \end{equation}
\end{itemize}

\section{Results and discussion}
\begin{widetext}
	we have studied the GmSUGRA	
\begin{figure}[h!]
	\centering \includegraphics[width=7.90cm]{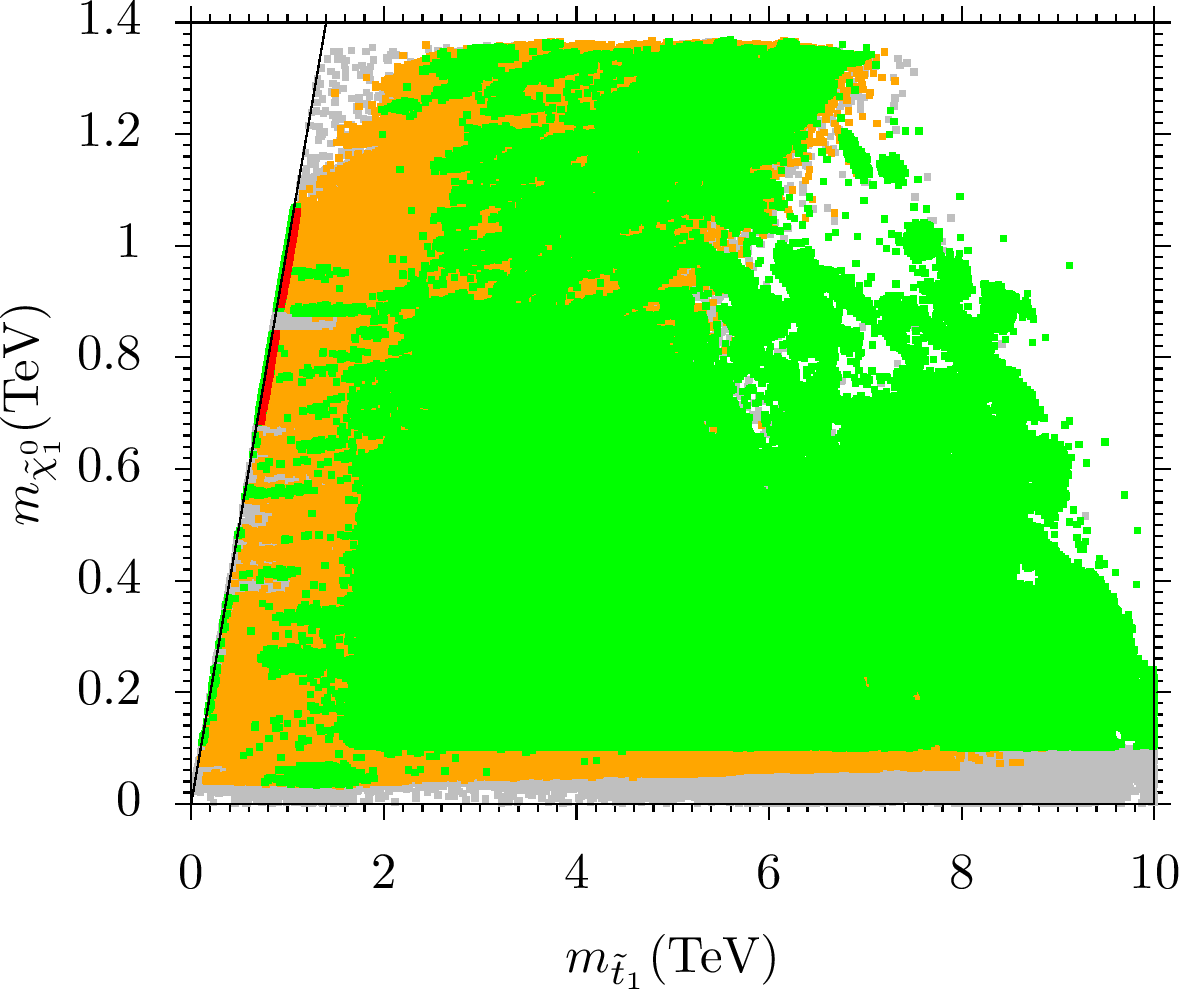}
	\centering \includegraphics[width=7.90cm]{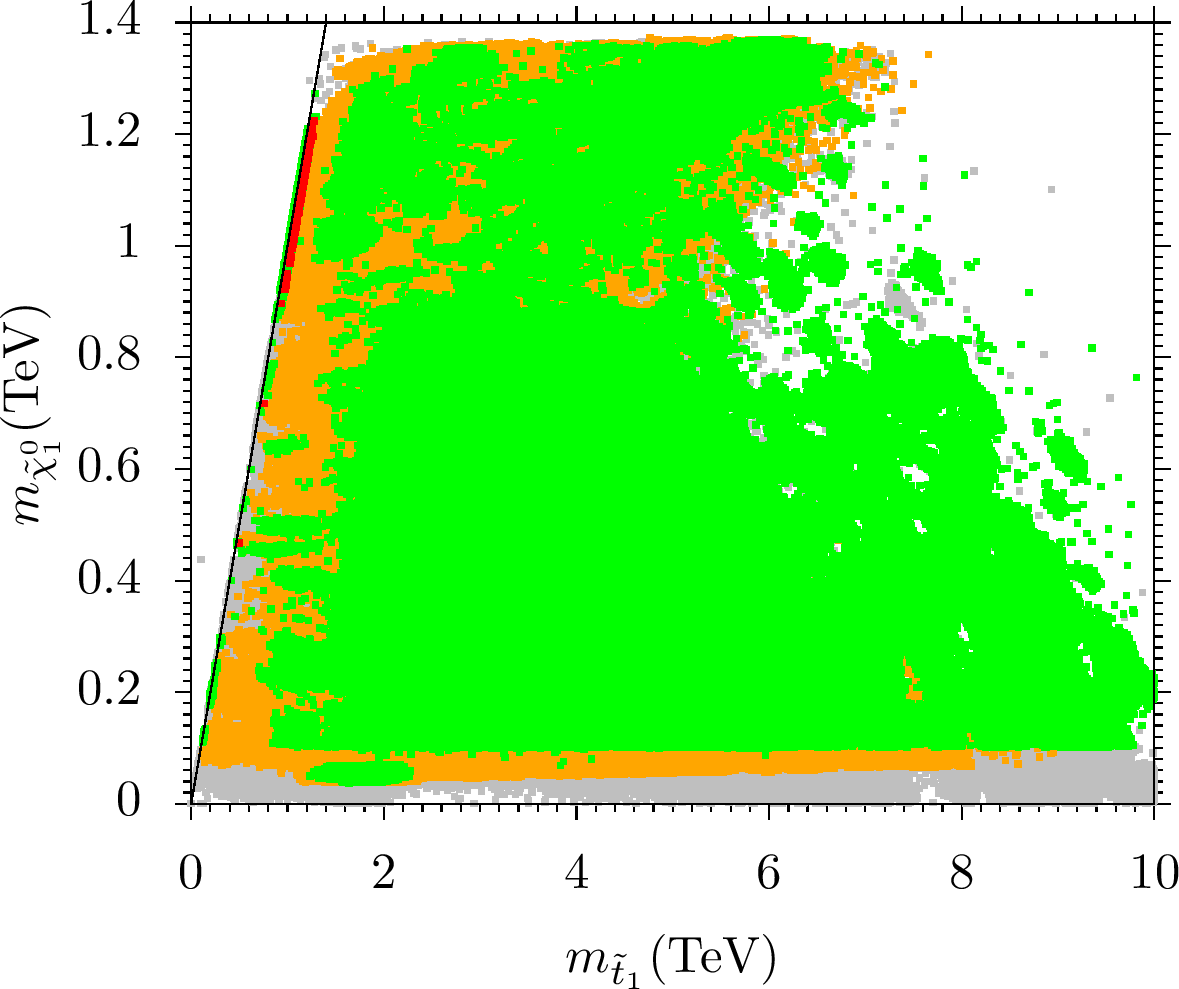}
	\centering \includegraphics[width=7.90cm]{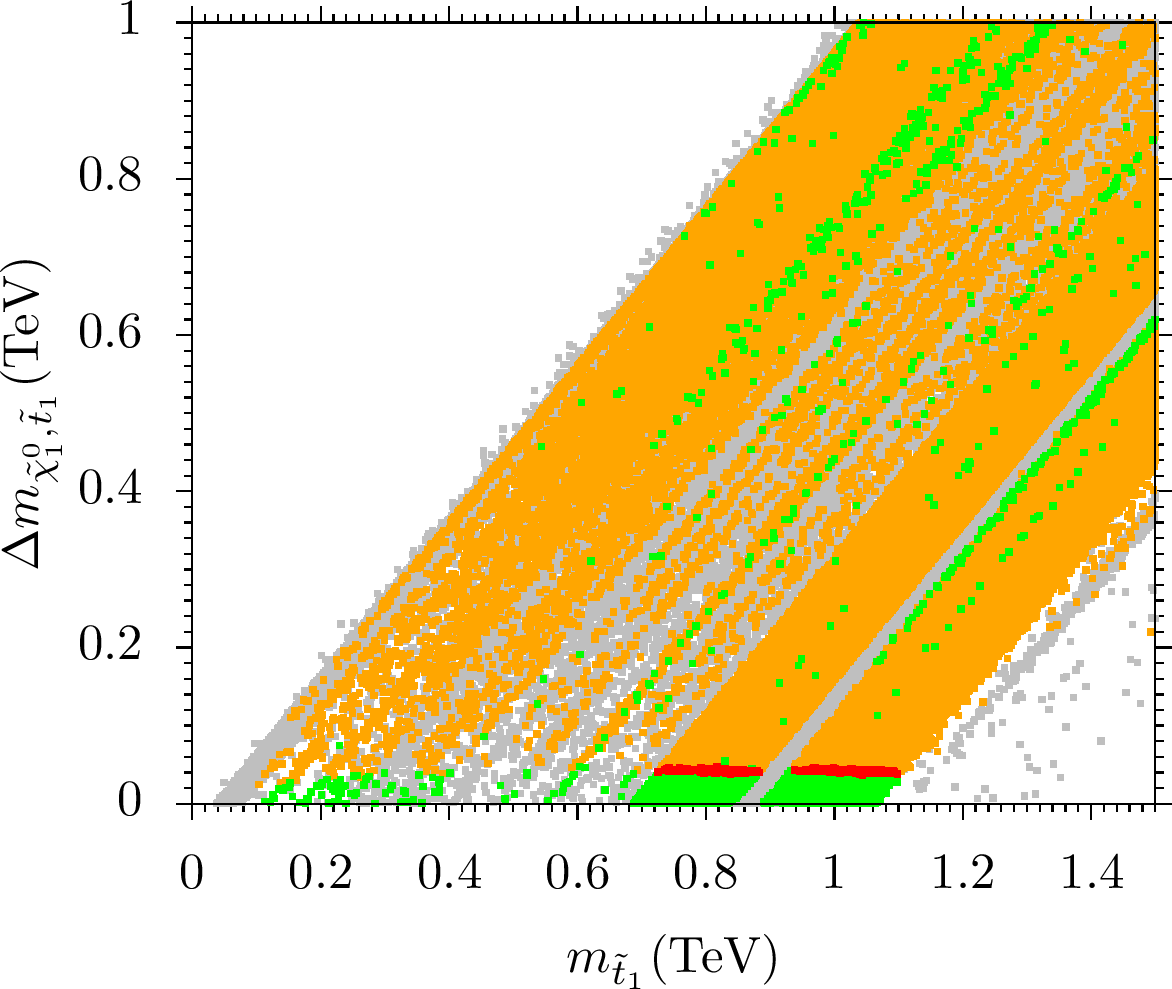}
	\centering \includegraphics[width=7.90cm]{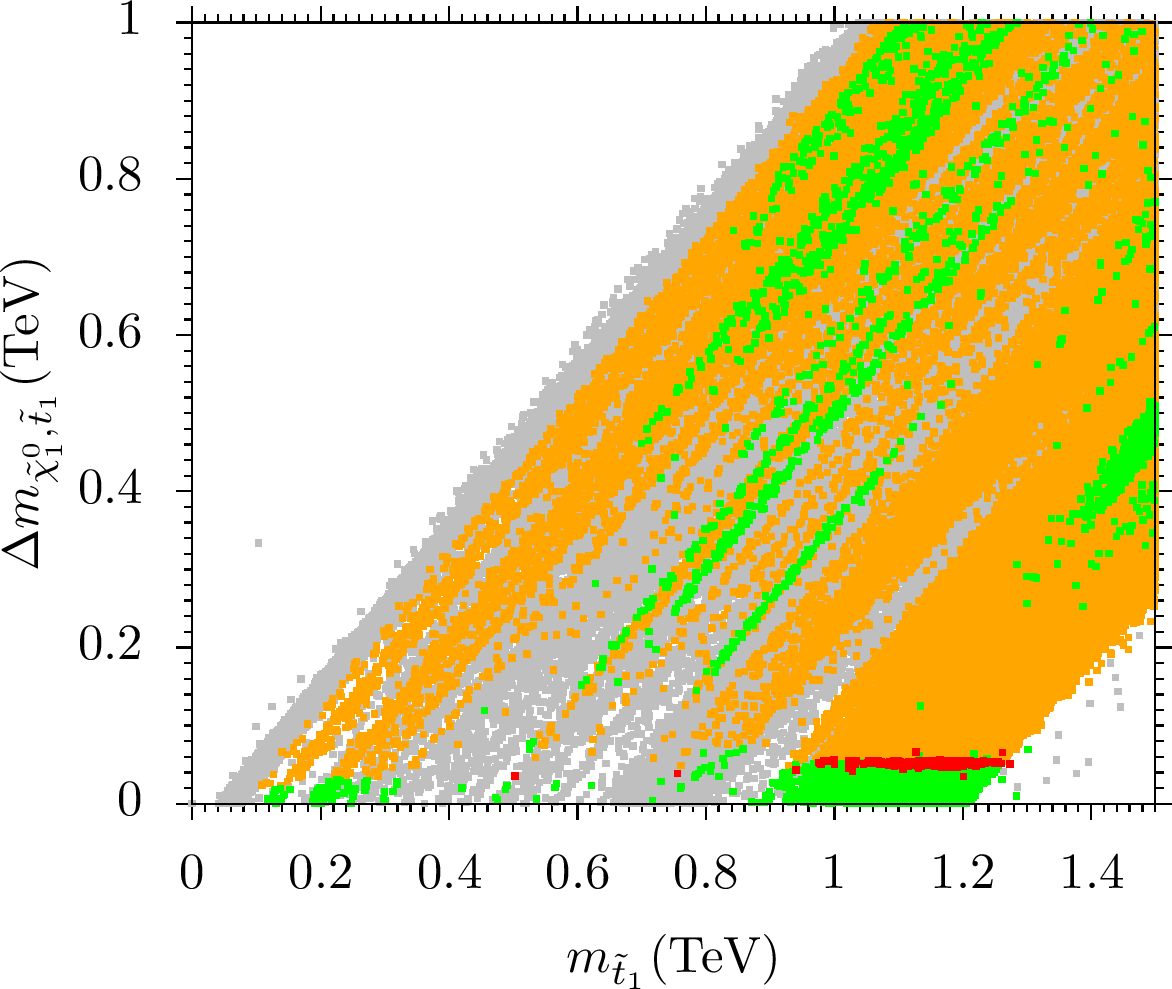}    
	\centering \includegraphics[width=7.90cm]{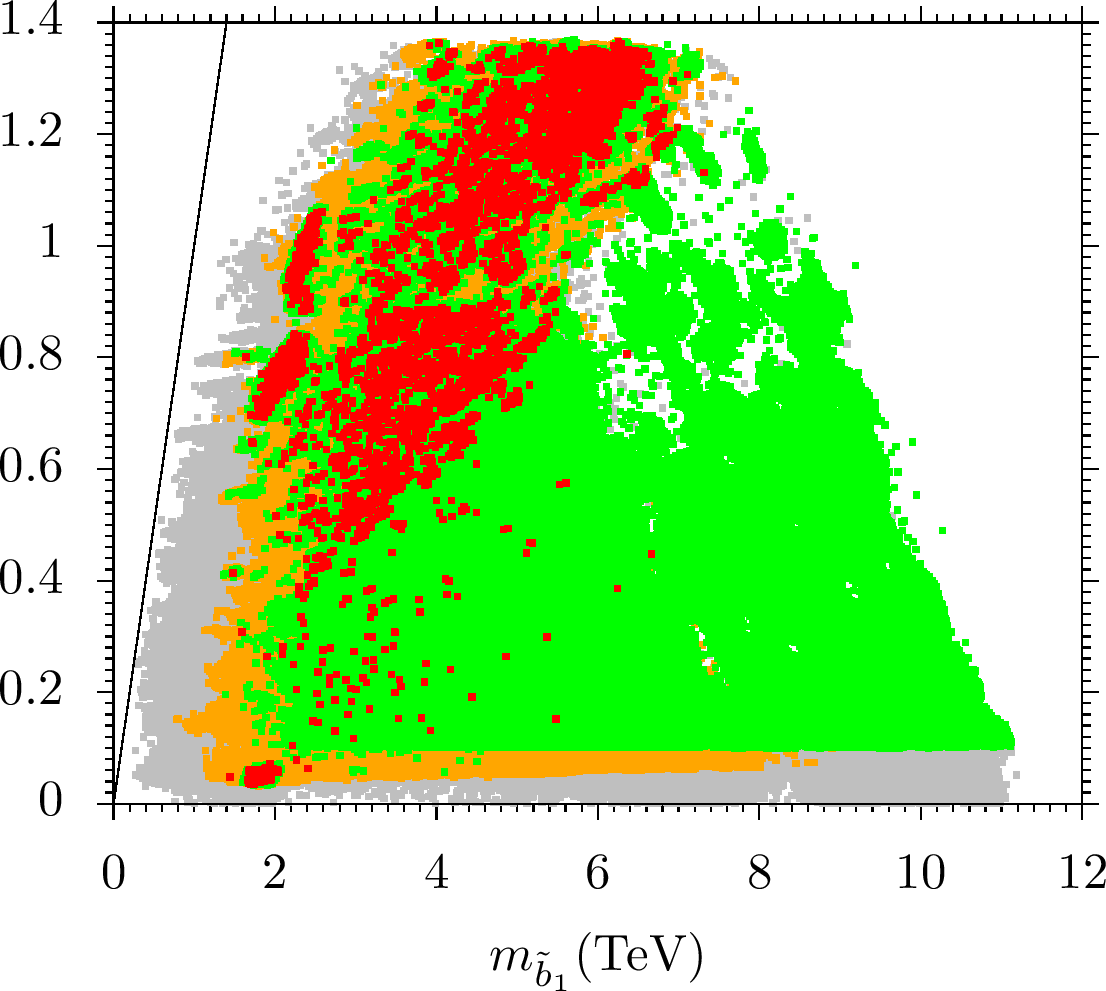}
	\centering \includegraphics[width=7.90cm]{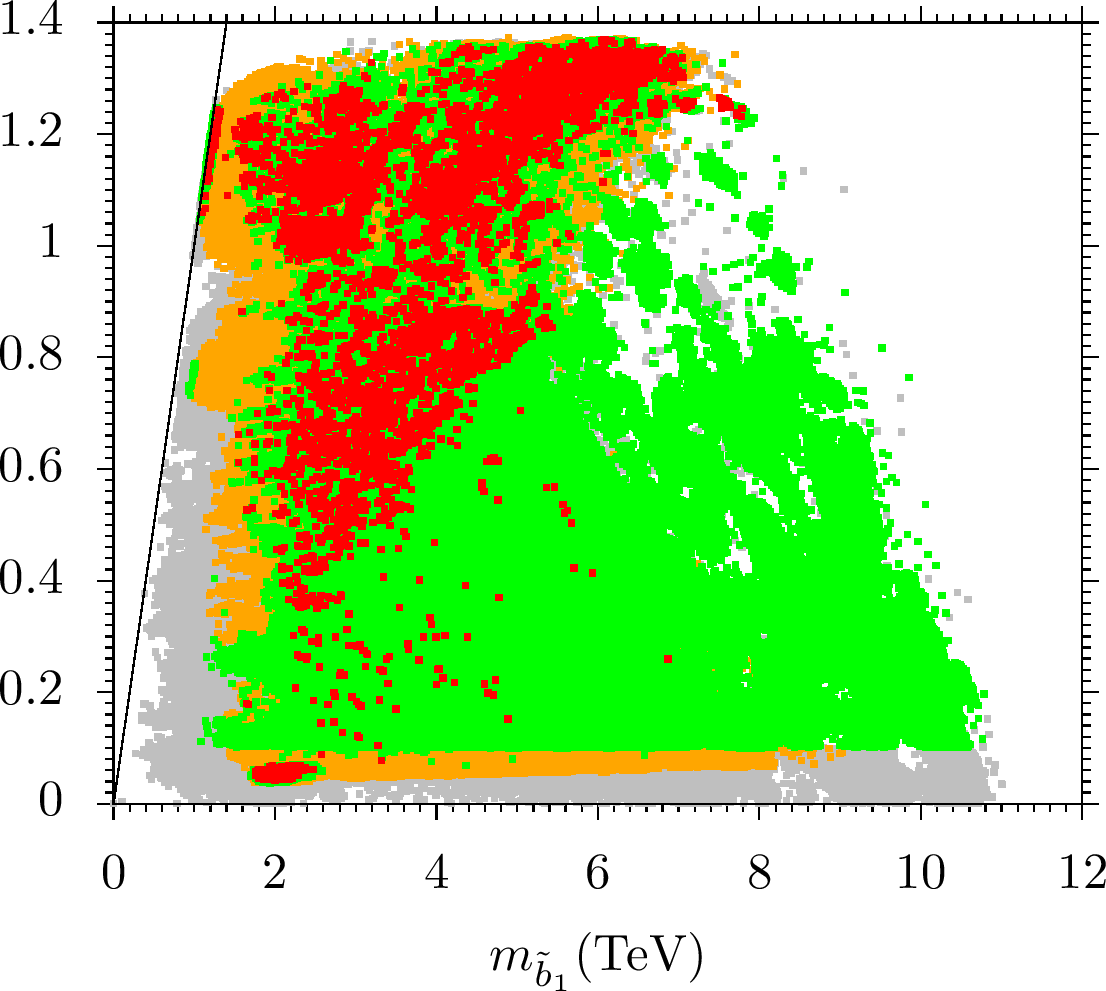}
\caption{\small$m_{\tilde t_1}$ vs. $m_{\tilde \chi_{1}^{0}}$, $m_{\tilde t_1}$ vs. $\vert \Delta m_{\tilde \chi_{1}^{0},\tilde t_{1}}\vert$, and $m_{\tilde b1}$ vs. $m_{\tilde \chi_{1}^{0}}$ planes with color coding as follows: The left and right panels correspond to the $\mu > 0$ and $\mu < 0$ scenarios, respectively. Grey points satisfy the REWSB and yield LSP neutralino. Orange points satisfy the LEP mass bound, B-physics bound, Higgs bound including S-particles LHC constraint and satisfy the oversaturated relic density bound. Green points are the subsets of orange points and satisfy underrsaturated relic density bounds. Red points are the subsets of green points that satisfy the Planck relic density bound. The solid, black lines serve as visual cues for coannihilation as well as resonance solutions' sensitivity.}
	\label{1}
\end{figure} 

\begin{figure}[h!]
\centering \includegraphics[width=7.90cm]{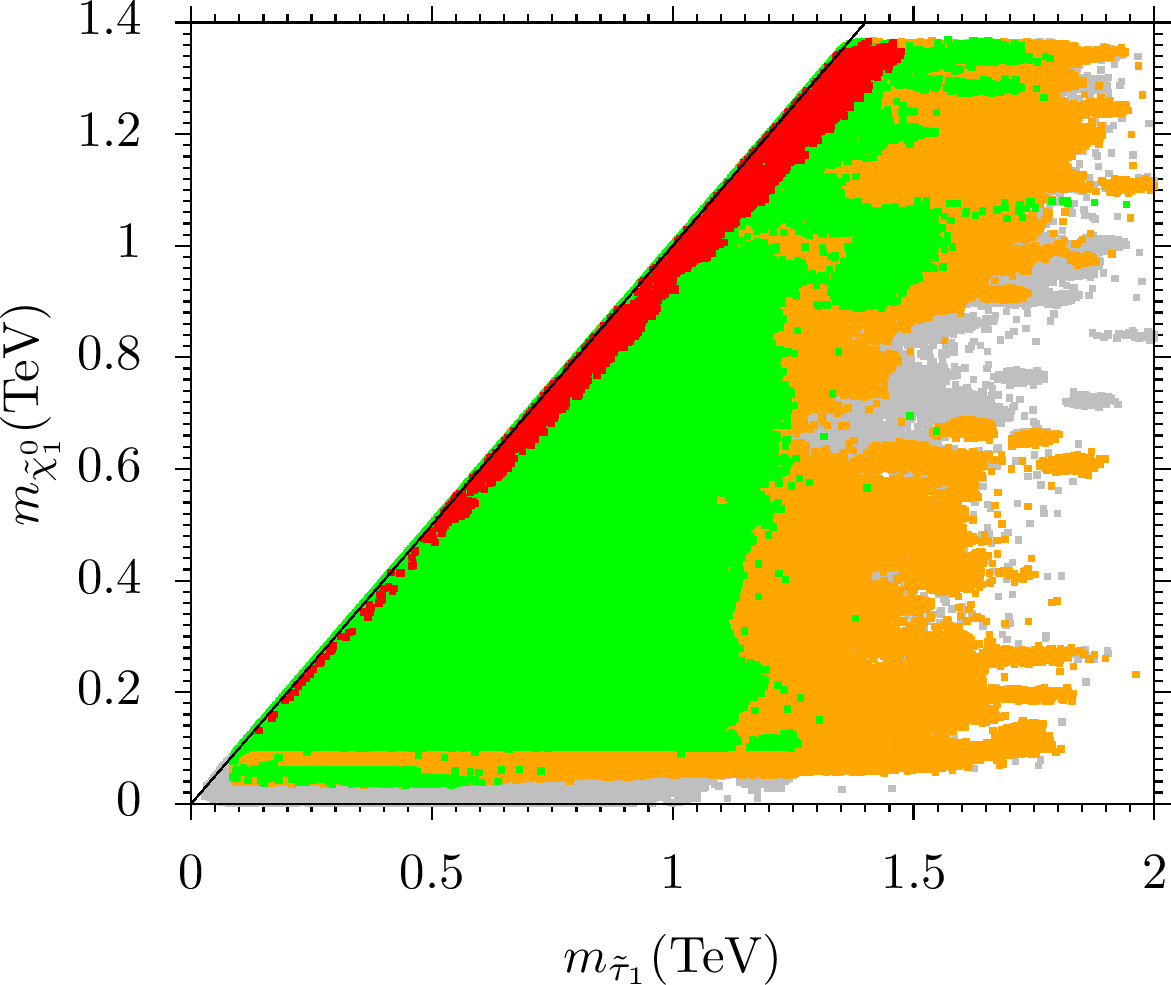}
	\centering \includegraphics[width=7.90cm]{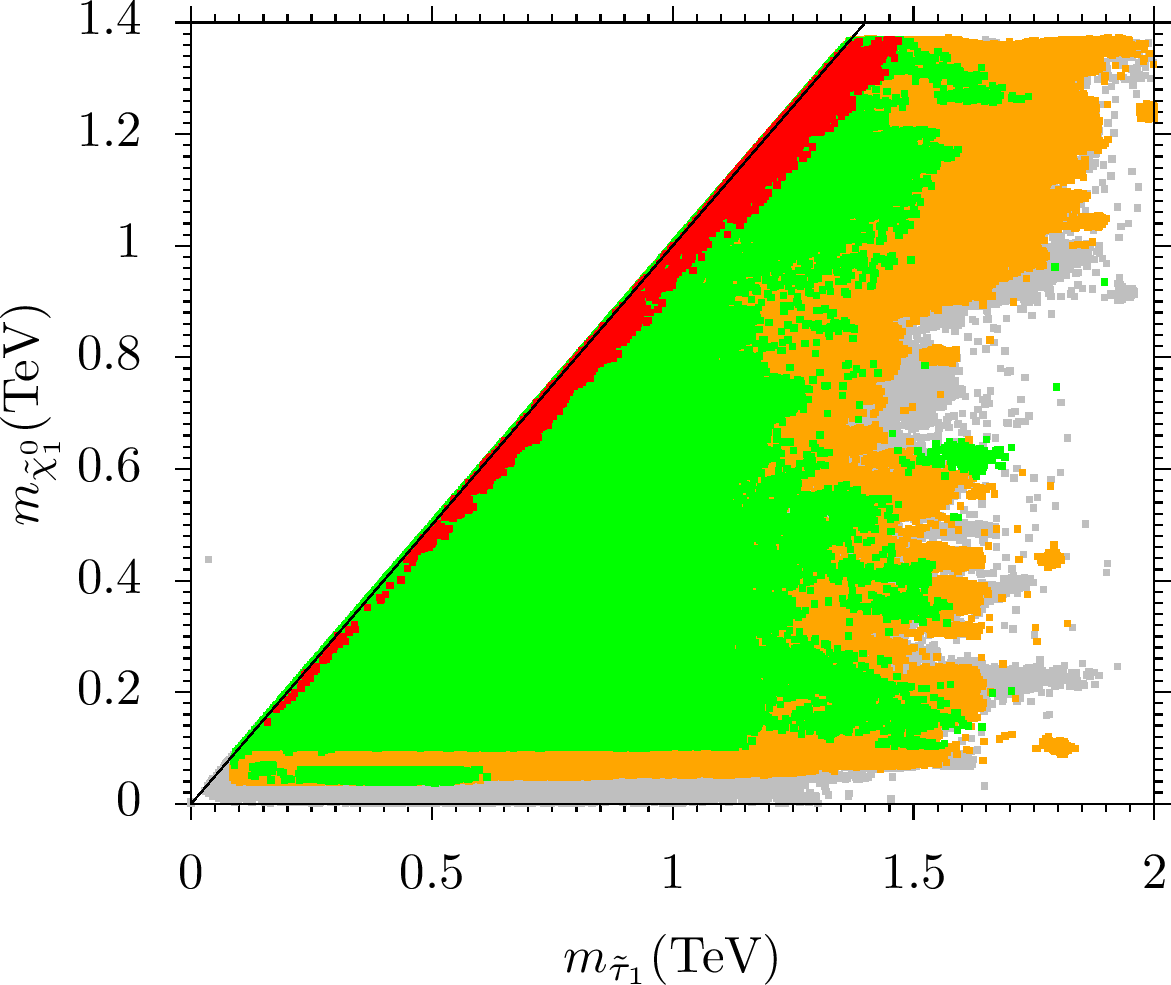}
	\centering \includegraphics[width=7.90cm]{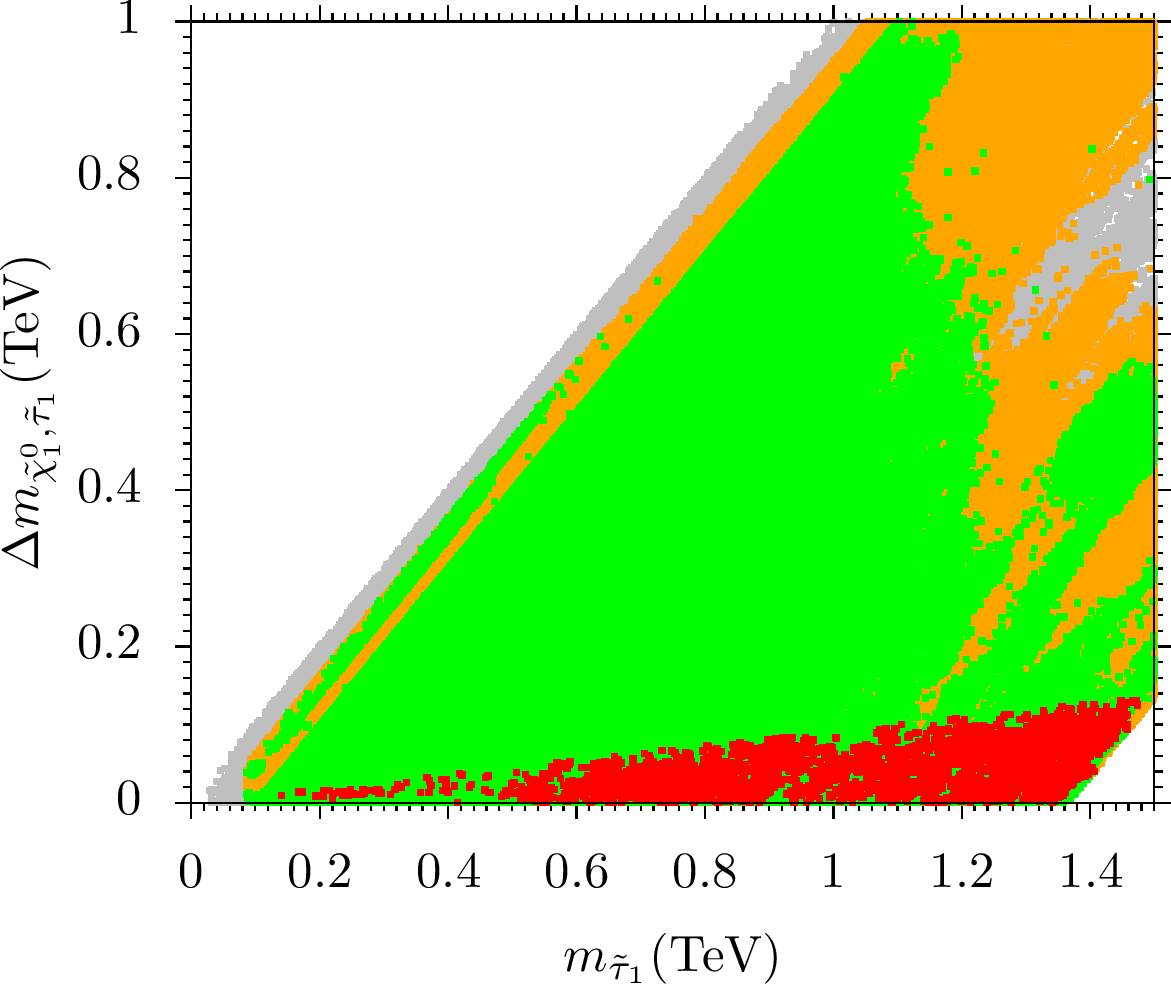}
	\centering \includegraphics[width=7.90cm]{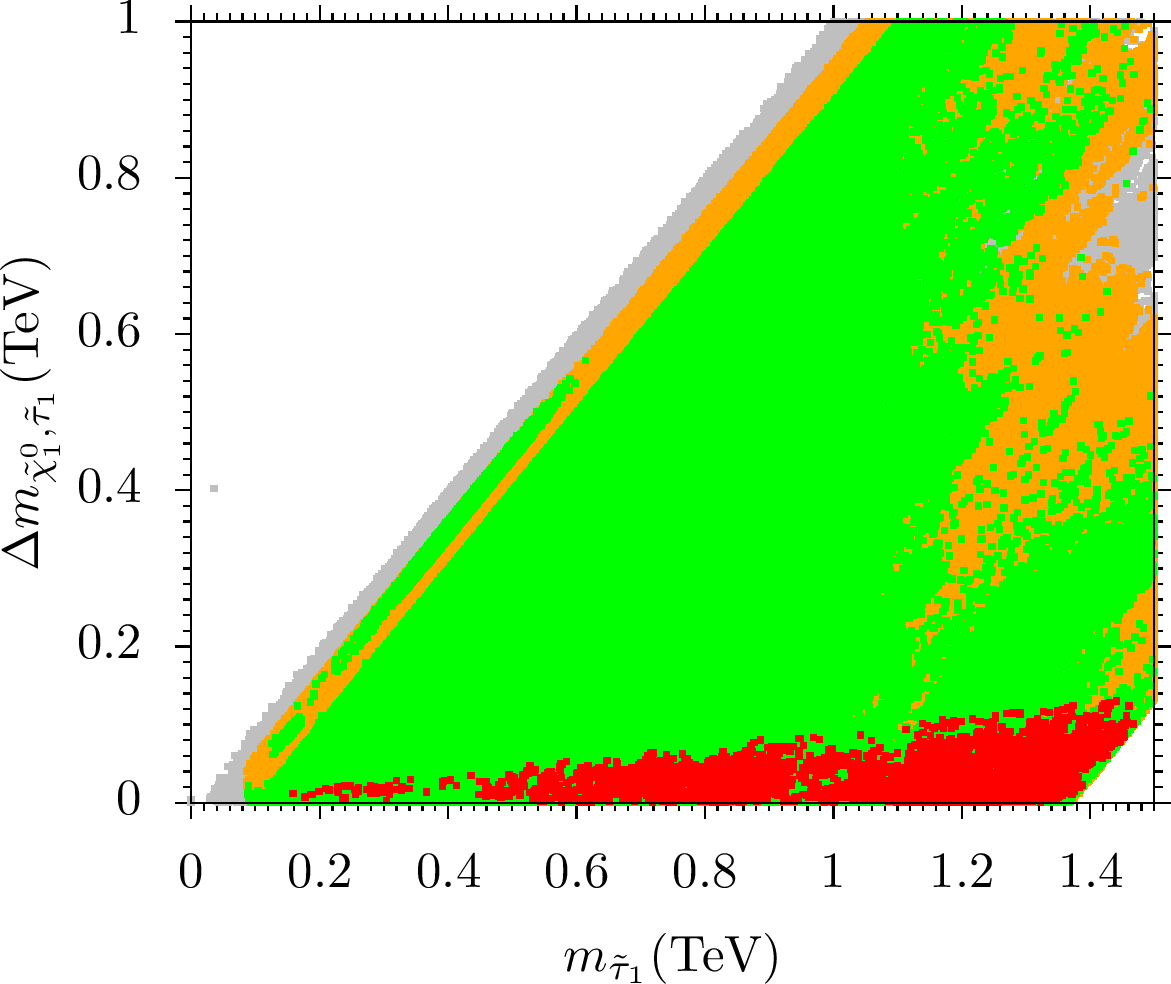}
	\centering \includegraphics[width=7.90cm]{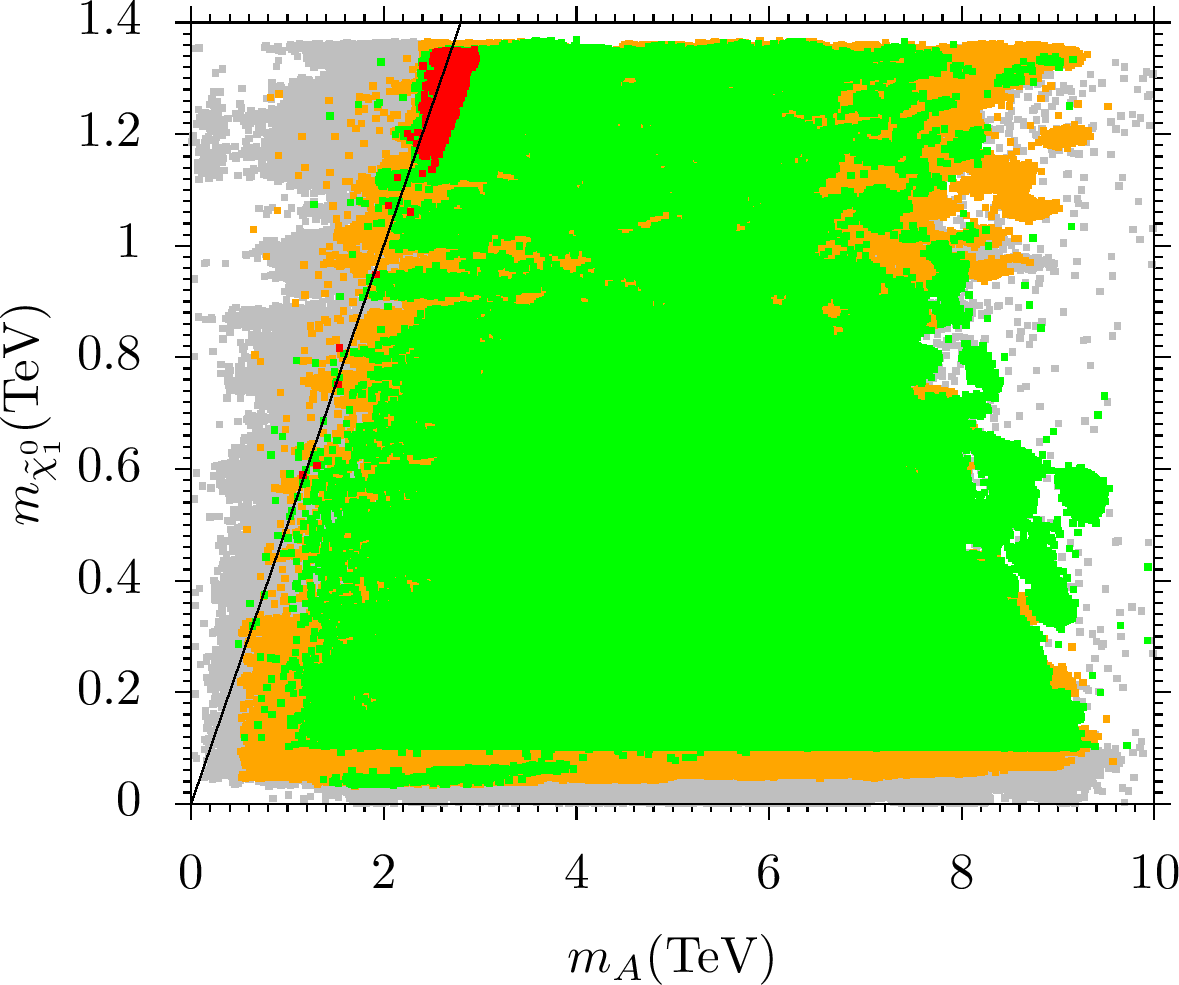}
	\centering \includegraphics[width=7.90cm]{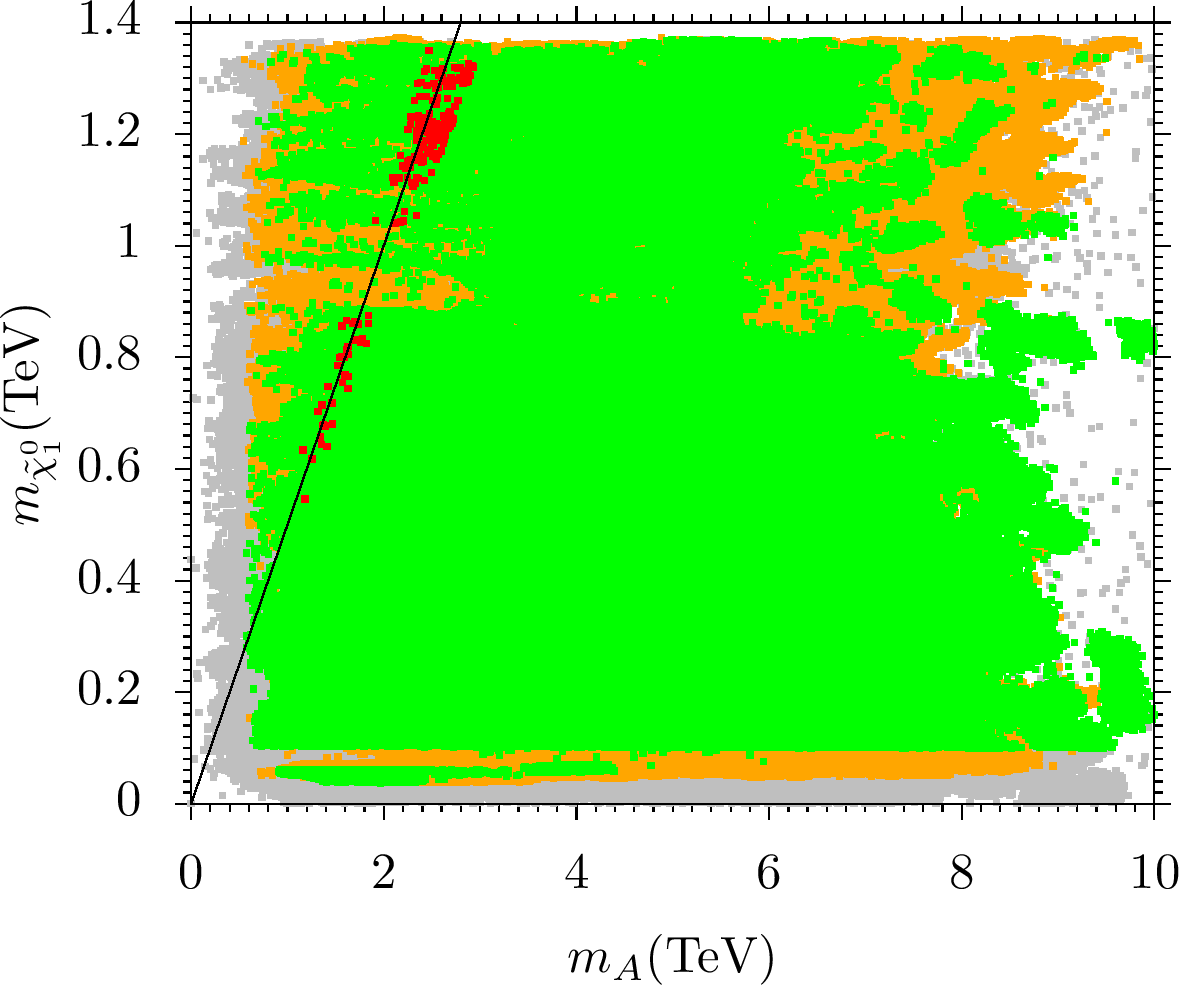}
	\caption{\small$m_{\tilde \tau_1}$ vs. $m_{\tilde \chi_{1}^{0}}$, $m_{\tilde \tau_{1}}$ vs. $\vert \Delta m_{\tilde \chi_{1}^{0},\tilde \tau_{1}}\vert$ and $m_{{A}}$ vs. $m_{\tilde \chi_{1}^{0}}$ planes with color coding as follows: The left and right panels correspond to the $\mu > 0$ and $\mu < 0$ scenarios, respectively. Grey points satisfy the REWSB and yield LSP neutralino. Orange points satisfy the LEP mass bound, B-physics bound, Higgs bound including S-particles LHC constraint, and satisfy the oversaturated relic density bound. Green points are the subsets of orange points and satisfy underrsaturated relic density bounds. Red points are the subsets of green points that satisfy the Planck relic density bound. The solid, black lines serve as visual cues for coannihilation as well as resonance solutions' sensitivity.}
	\label{2}
\end{figure}

\begin{figure}[h!]
	\centering \includegraphics[width=7.90cm]{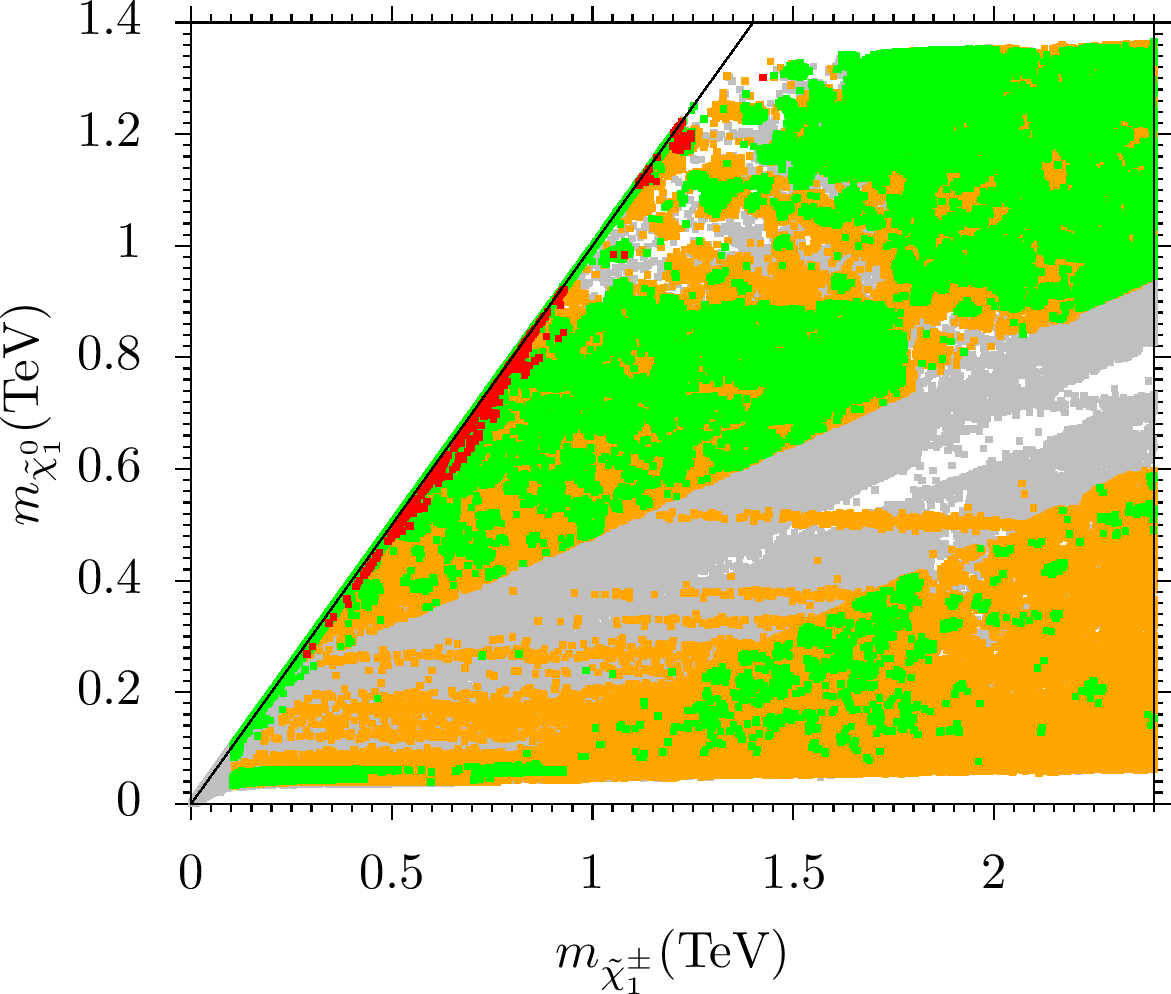}
	\centering \includegraphics[width=7.90cm]{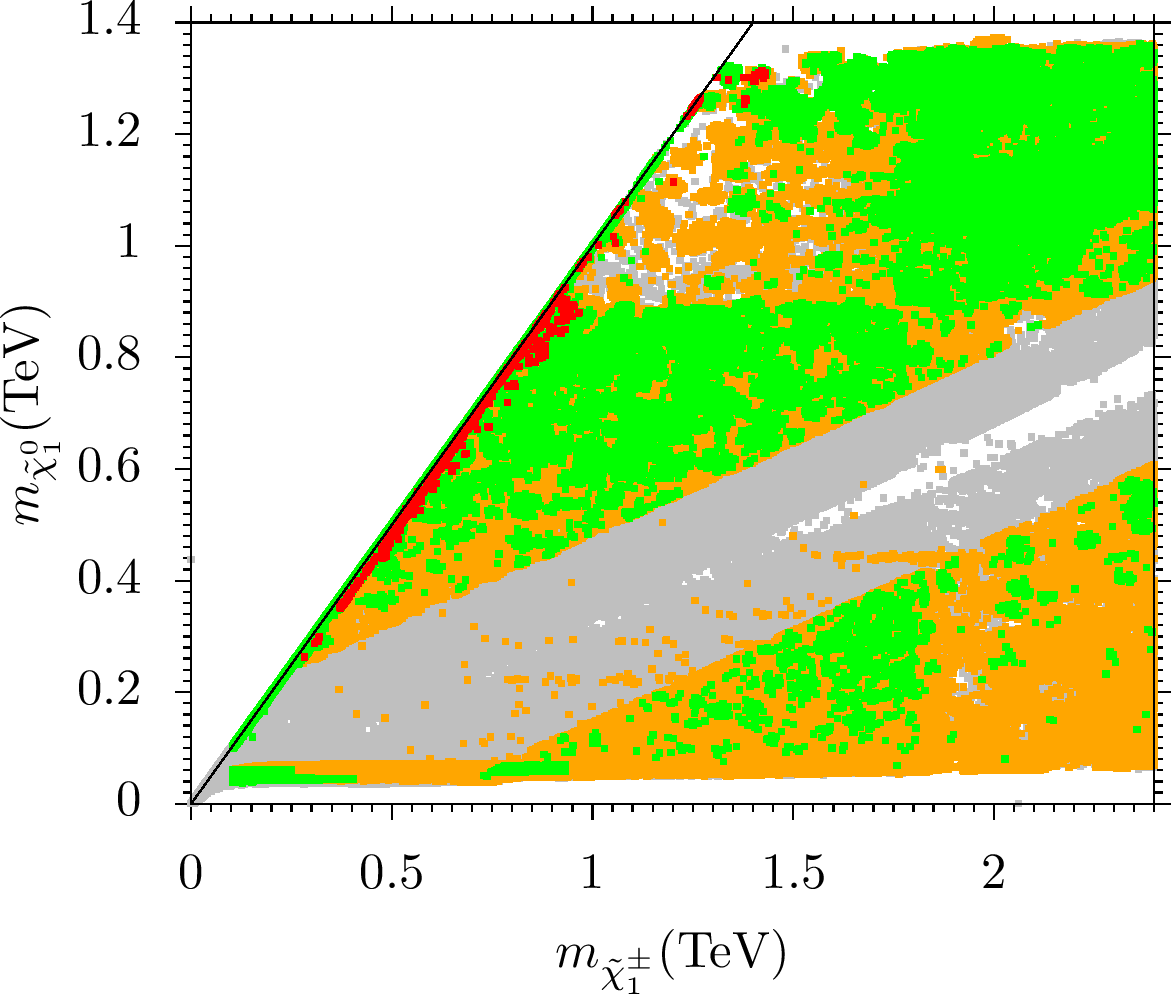}
    	\centering \includegraphics[width=7.90cm]{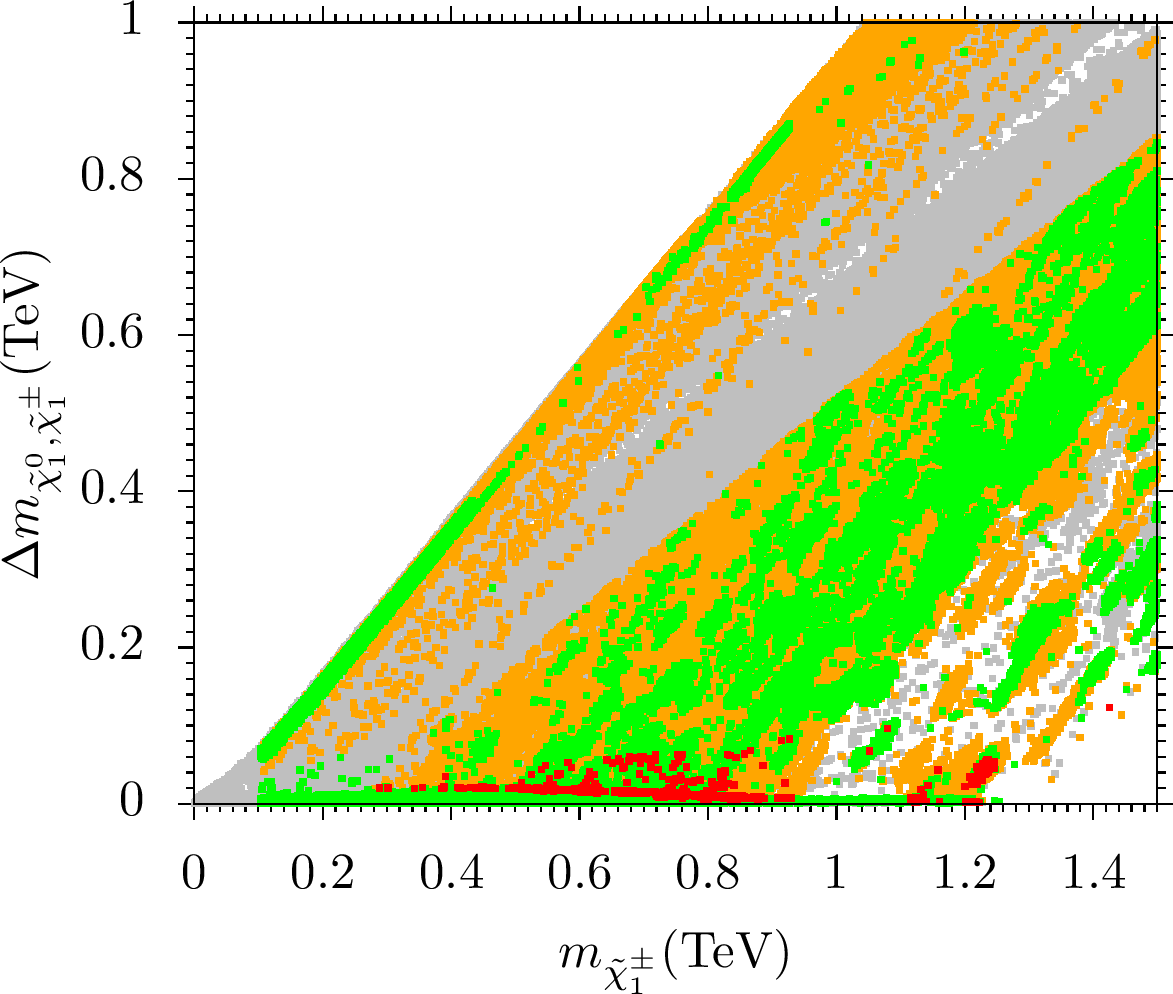}
	\centering \includegraphics[width=7.90cm]{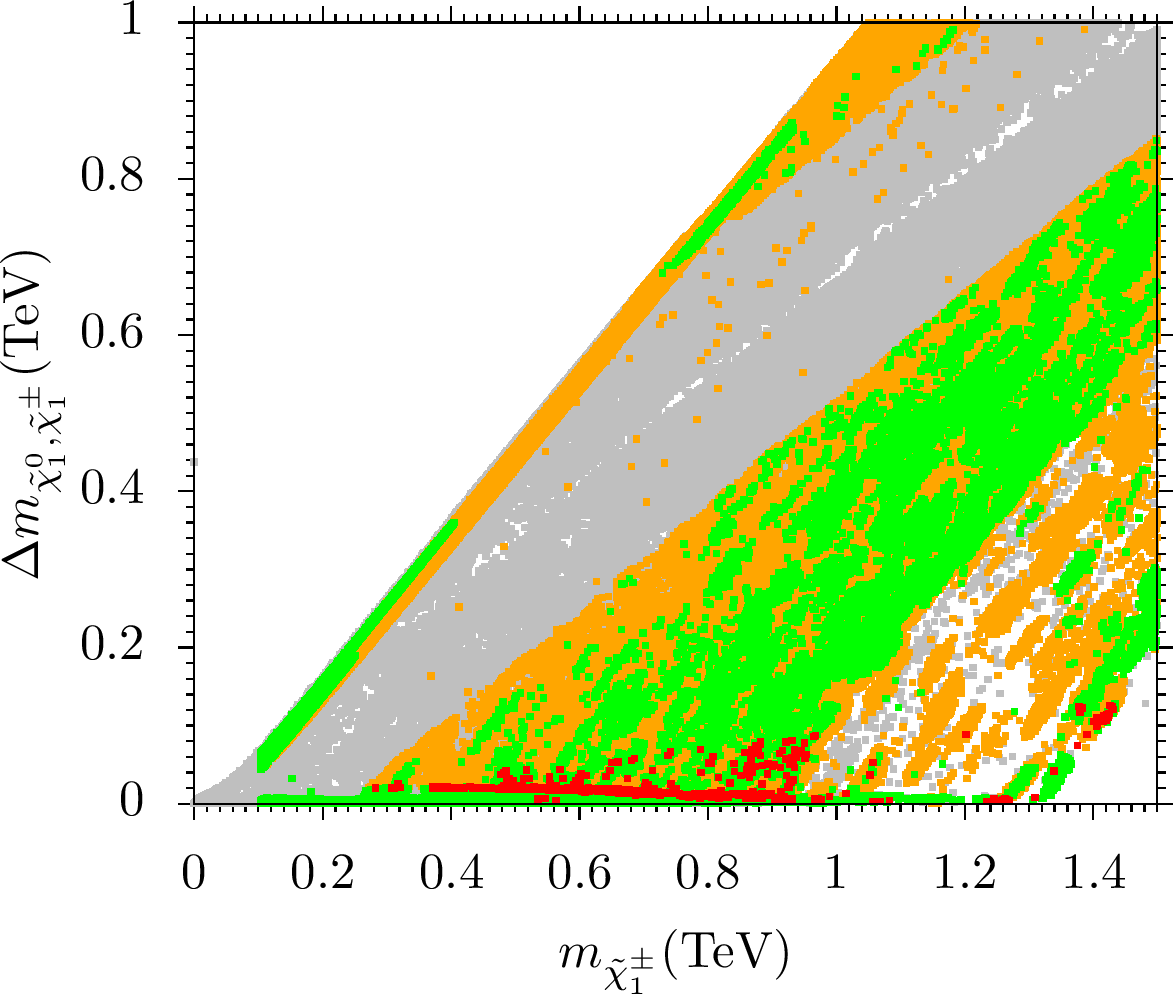}
	\centering \includegraphics[width=7.90cm, height=7.9cm]{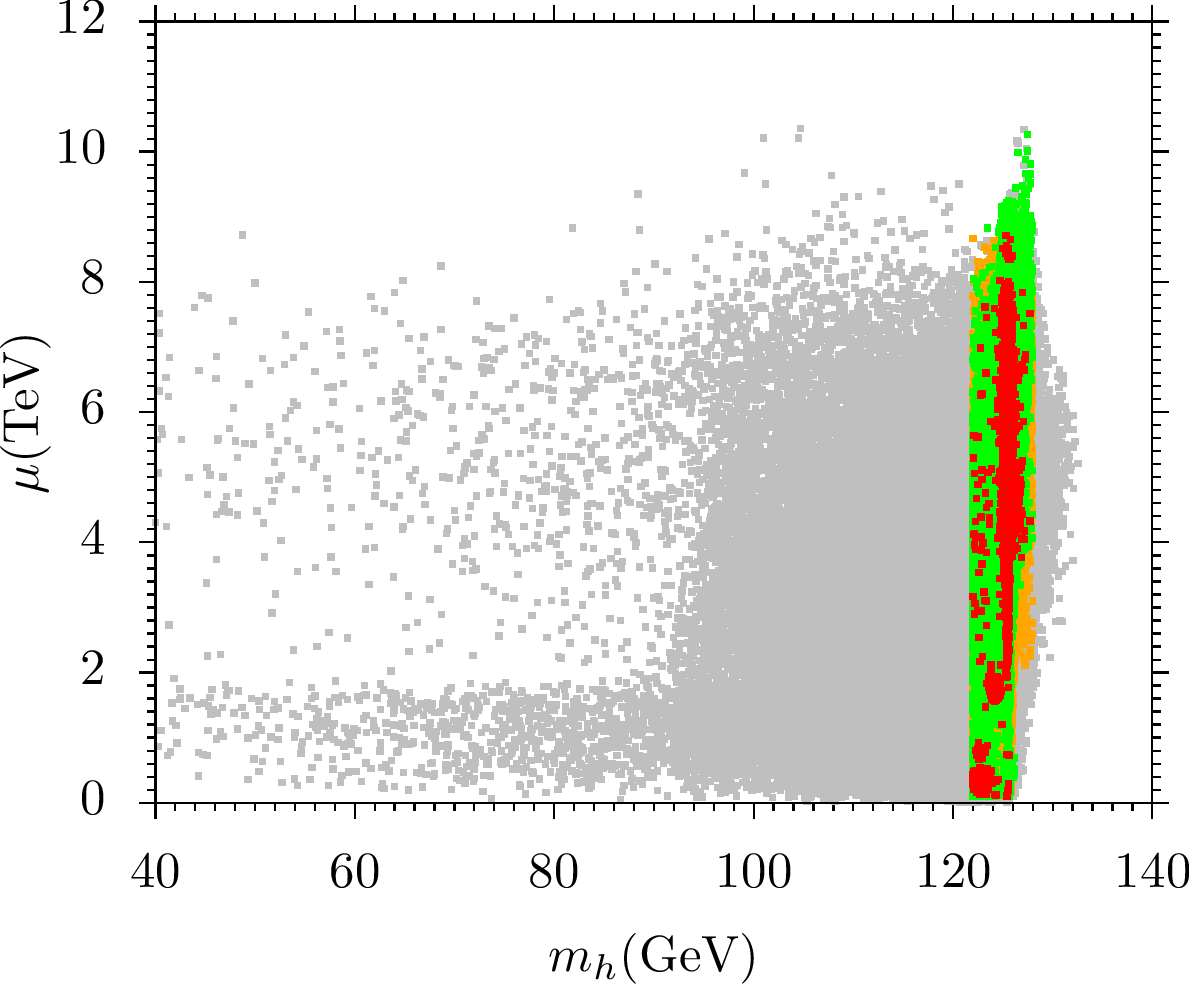}
	\centering \includegraphics[width=7.90cm,height=7.9cm]{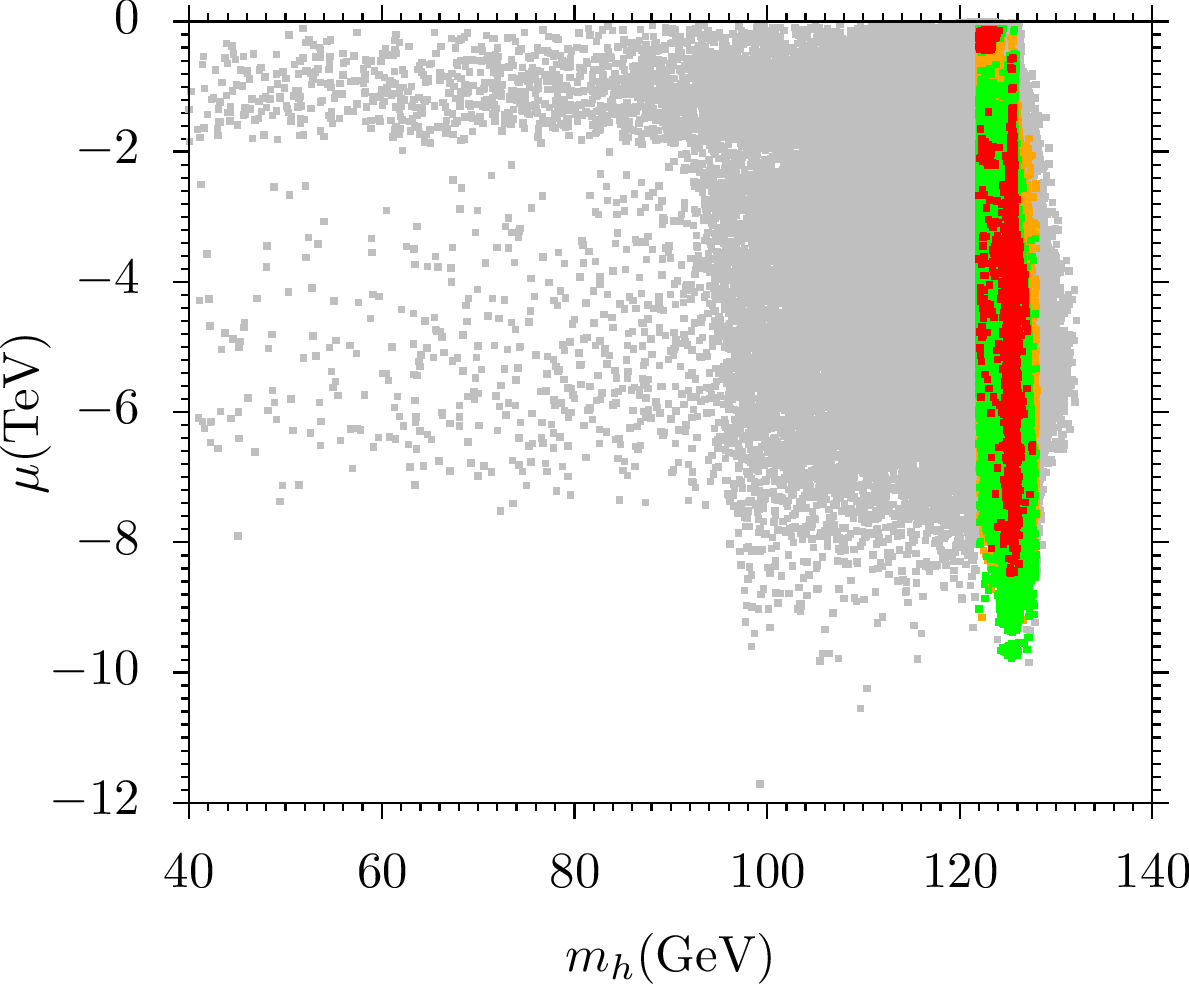}
	\caption{\small  $m_{\tilde \chi_{1}^{\pm}}$ vs. $m_{\tilde \chi_{0}^{\pm}}$, $m_{\tilde \chi_{1}^{\pm}}$ vs. $\mid \Delta m_{\tilde \chi_{1}^{0},\tilde \chi_{1}^{\pm}}\mid$, and 
 $\mu$ vs. $m_{{h}}$ planes with color coding as follows: The left and right panels correspond to the $\mu > 0$ and $\mu < 0$ scenarios, respectively. Grey points satisfy the REWSB and yield LSP neutralino. Orange points satisfy the LEP mass bound, B-physics bound, Higgs bound including S-particles LHC constraint, and satisfy the oversaturated relic density bound. Green points are the subsets of orange points and satisfy underrsaturated relic density bounds. Red points are the subsets of green points that satisfy the Planck relic density bound. The solid, black lines serve as visual cues for coannihilation as well as resonance solutions' sensitivity.}
	\label{3}
\end{figure} 
\begin{figure}[h!]
\centering \includegraphics[width=7.90cm]{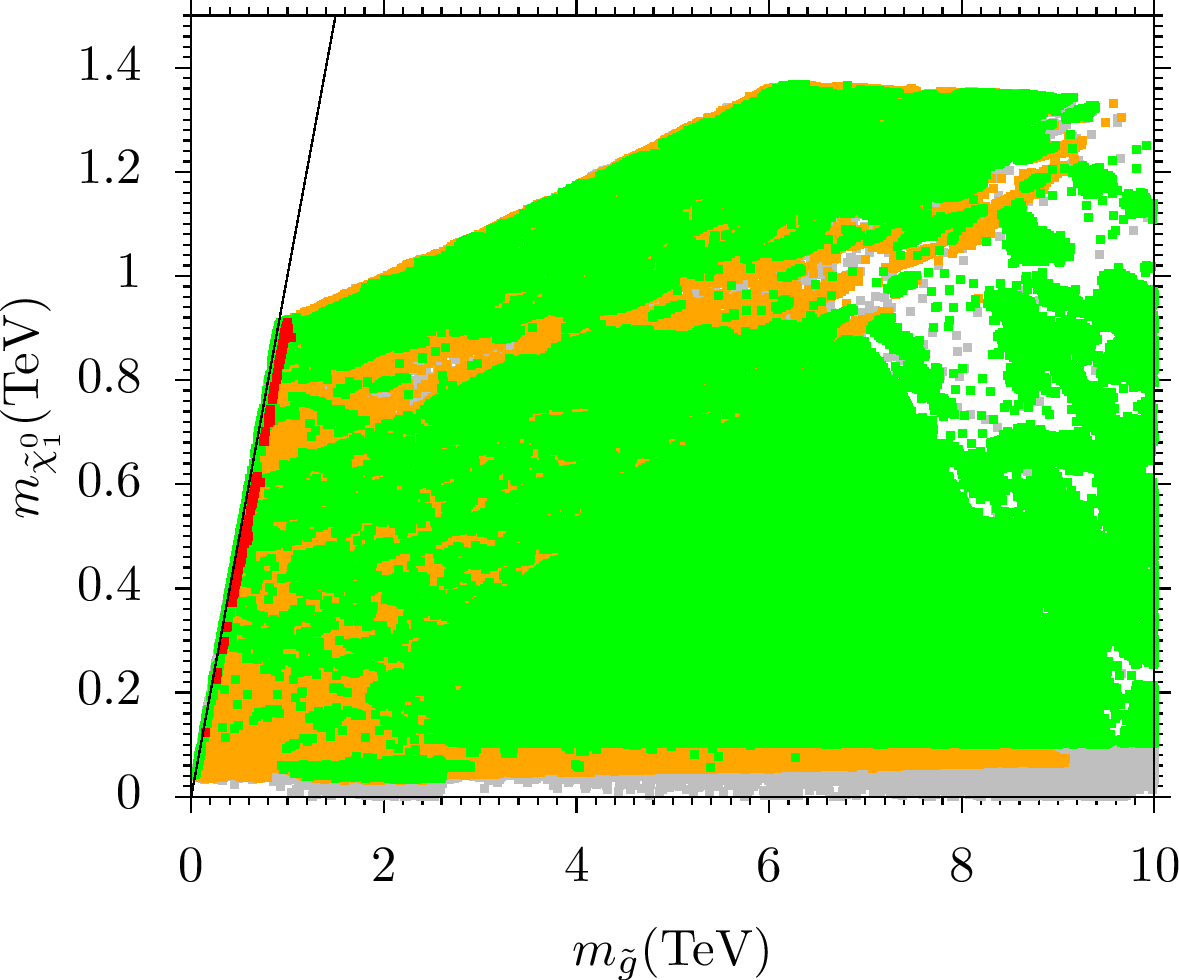}
	\centering \includegraphics[width=7.90cm]{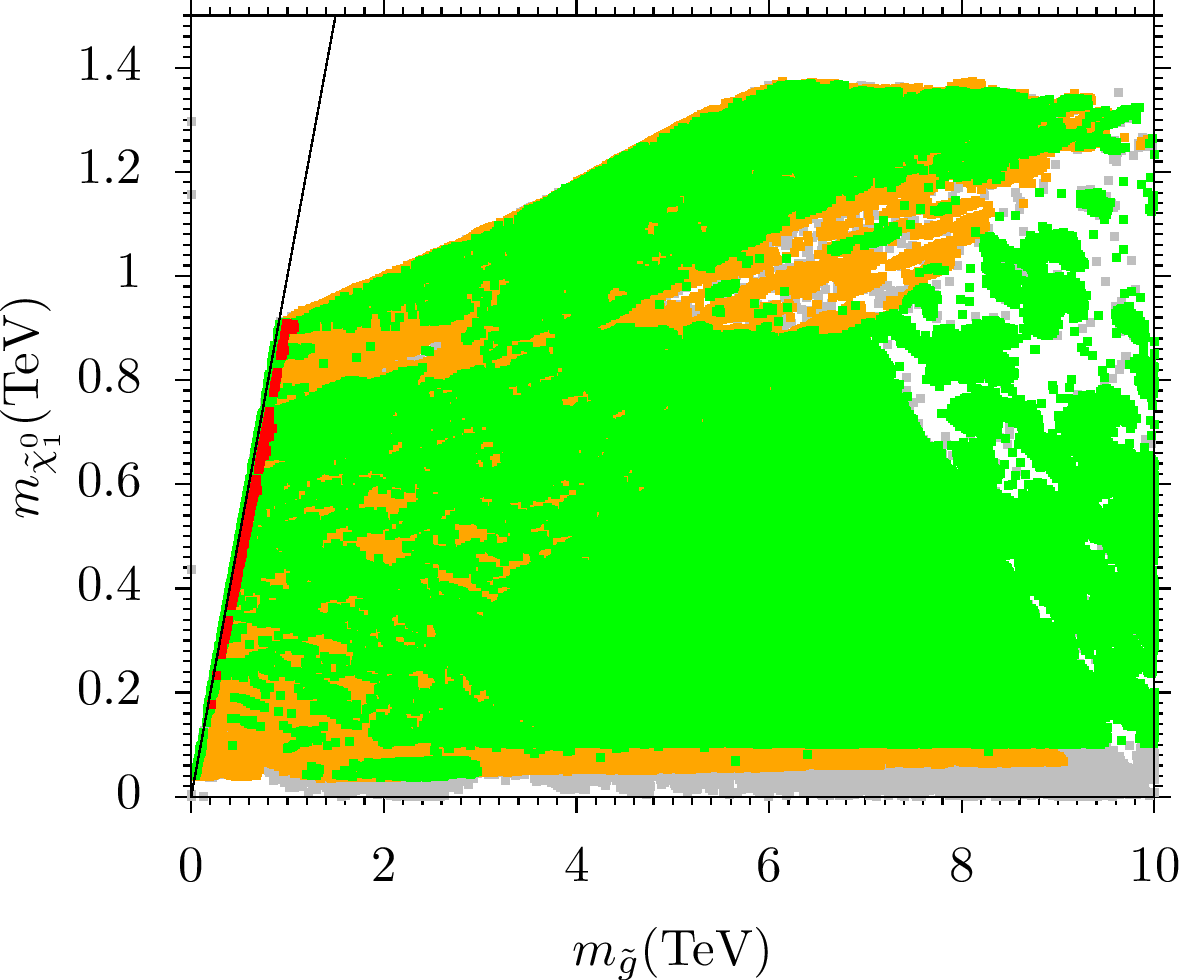}
	\centering \includegraphics[width=7.90cm]{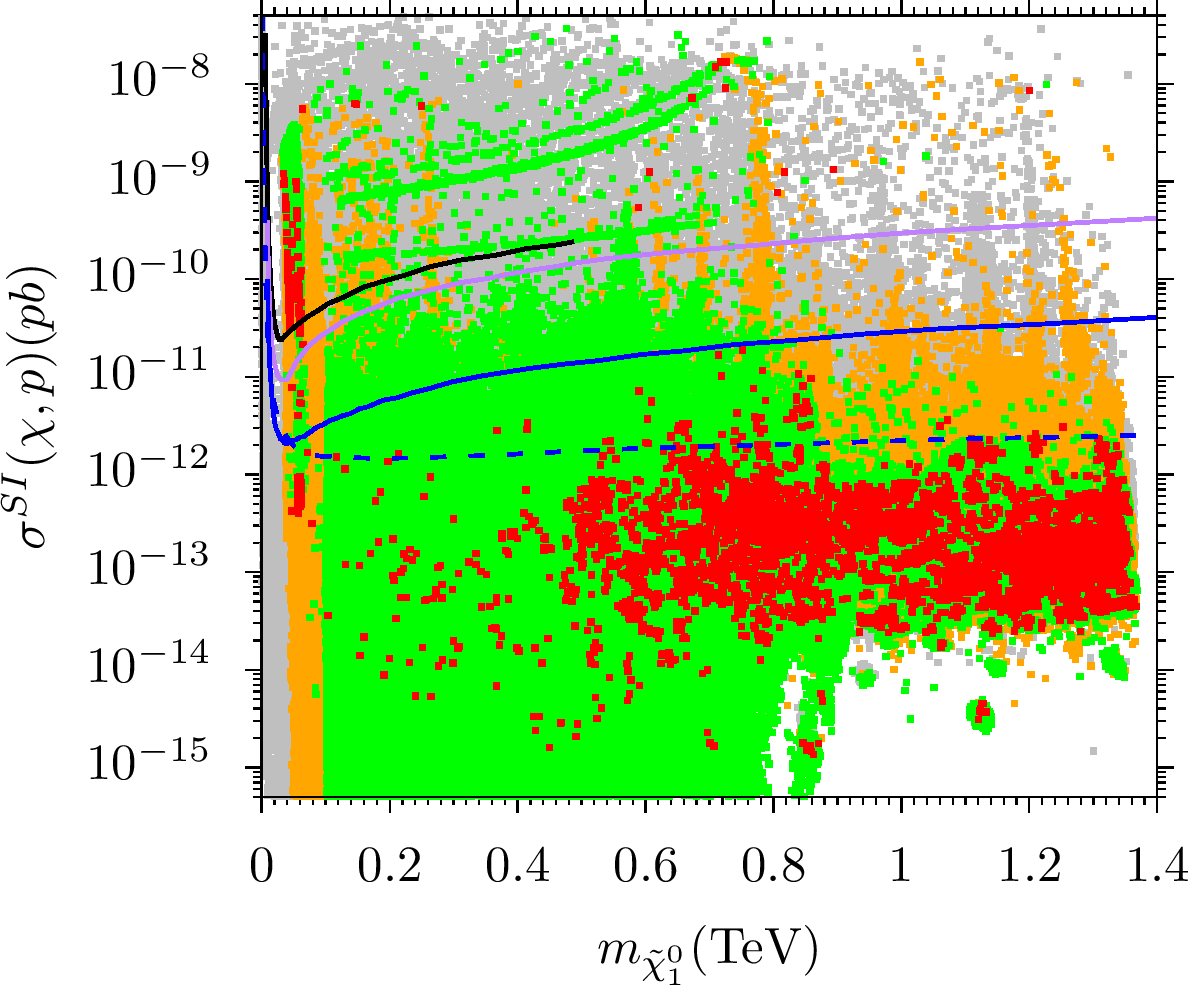}
	\centering \includegraphics[width=7.90cm]{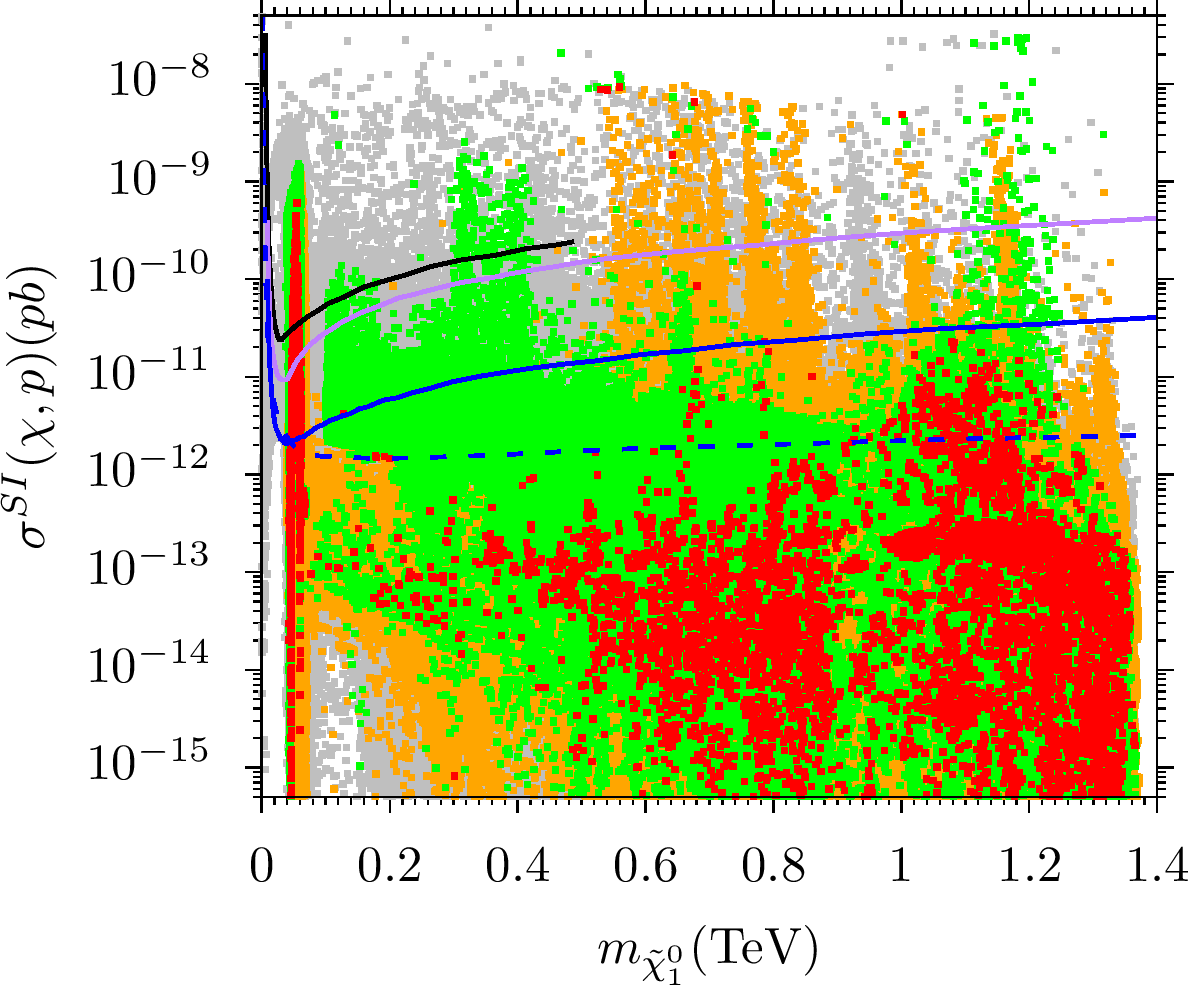}
	\centering \includegraphics[width=7.90cm]{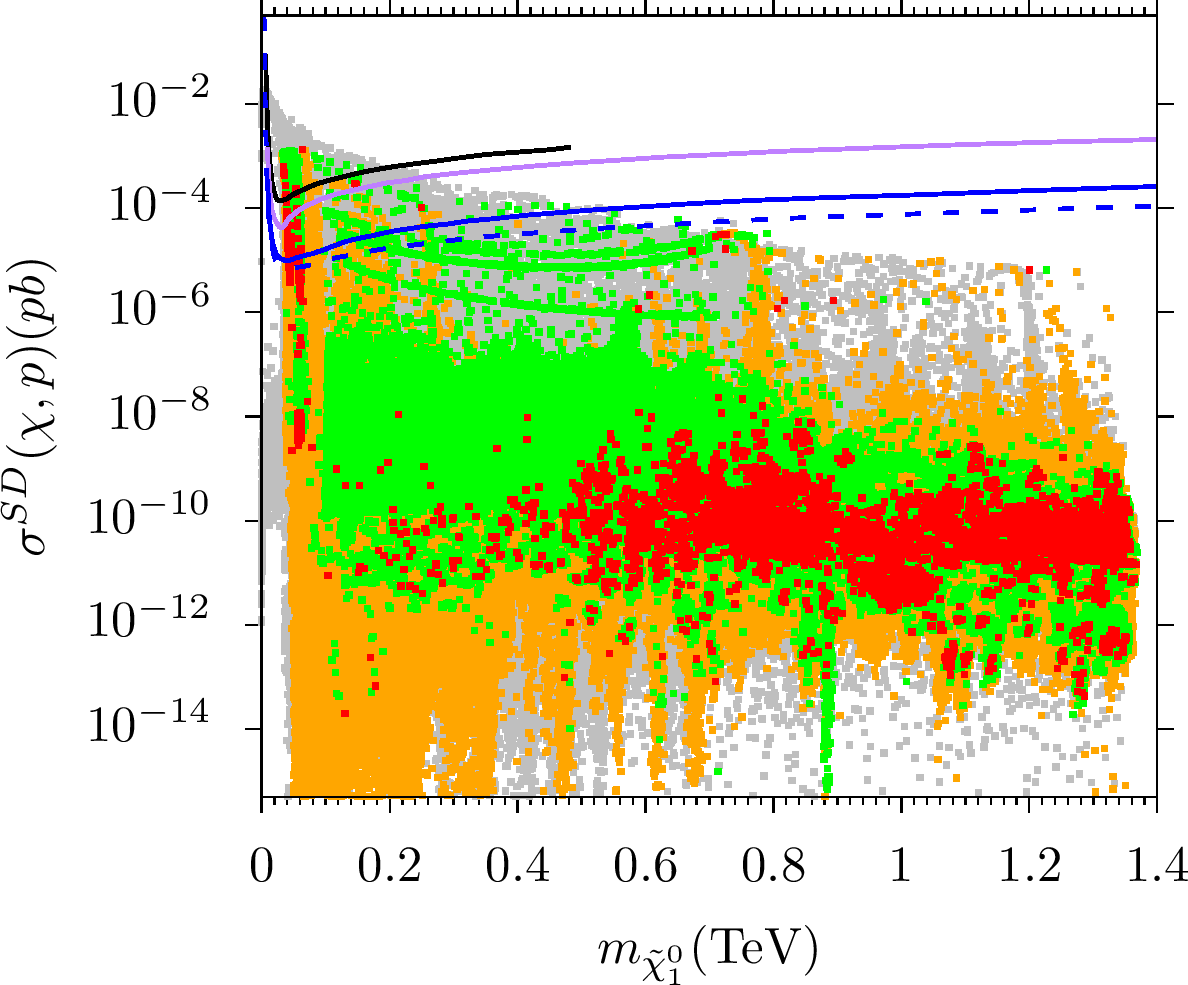}
	\centering \includegraphics[width=7.90cm]{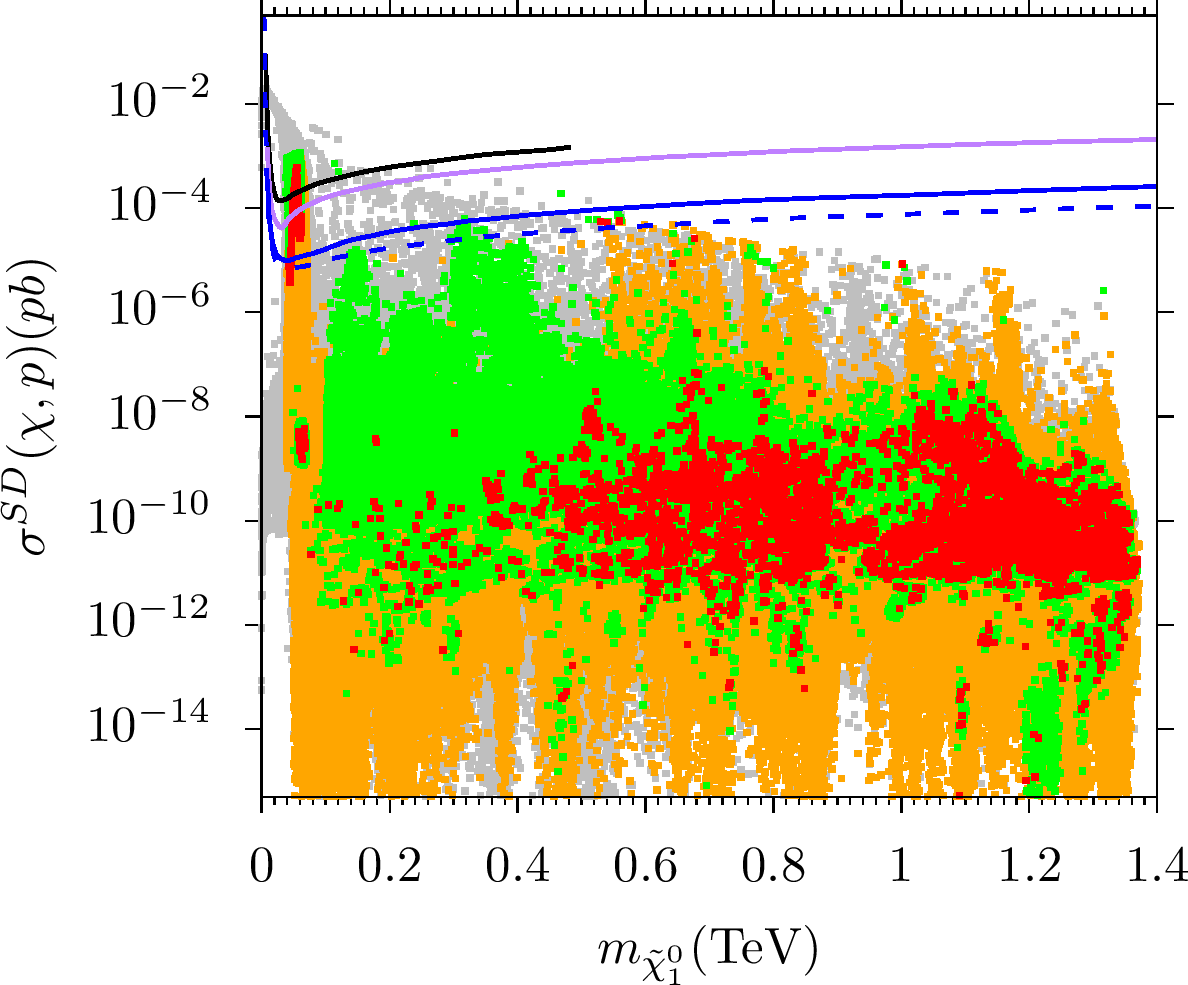}
	\caption{\small $m_{\tilde{\chi}^0_{1}}$ vs. $m_{\tilde {g}}$, $\sigma^{SI}(\chi,p)(pb)$ and $\sigma^{SD}(\chi,p)(pb)$: The left and right panels correspond to the $\mu > 0$ and $\mu < 0$ scenarios, respectively. Grey points satisfy the REWSB, resulting in LSP neutralino.  Orange points satisfy the LEP mass bounds, B-physics bounds, Higgs bound including S-particles LHC constraints, and satisfy the oversaturated relic density bound. Green points are the subsets of orange points and satisfy underrsaturated relic density bounds. Red points are the subsets of green points that satisfy  the Planck relic density bound. The solid black line represents the XENONnT \cite{XENON:2023cxc} and for the LUX-ZEPLIN, it is as follows: the Solid purple line represents LZ (2022), the solid blue line represents LZ (2024), and the dotted blue line represents the LZ-1000 day sensitivity~\cite{LZ:2018qzl, LZ:2022lsv,LZ:2024zvo}}
	\label{4}
\end{figure} 

\begin{figure}[h!]
\centering \includegraphics[width=7.90cm]{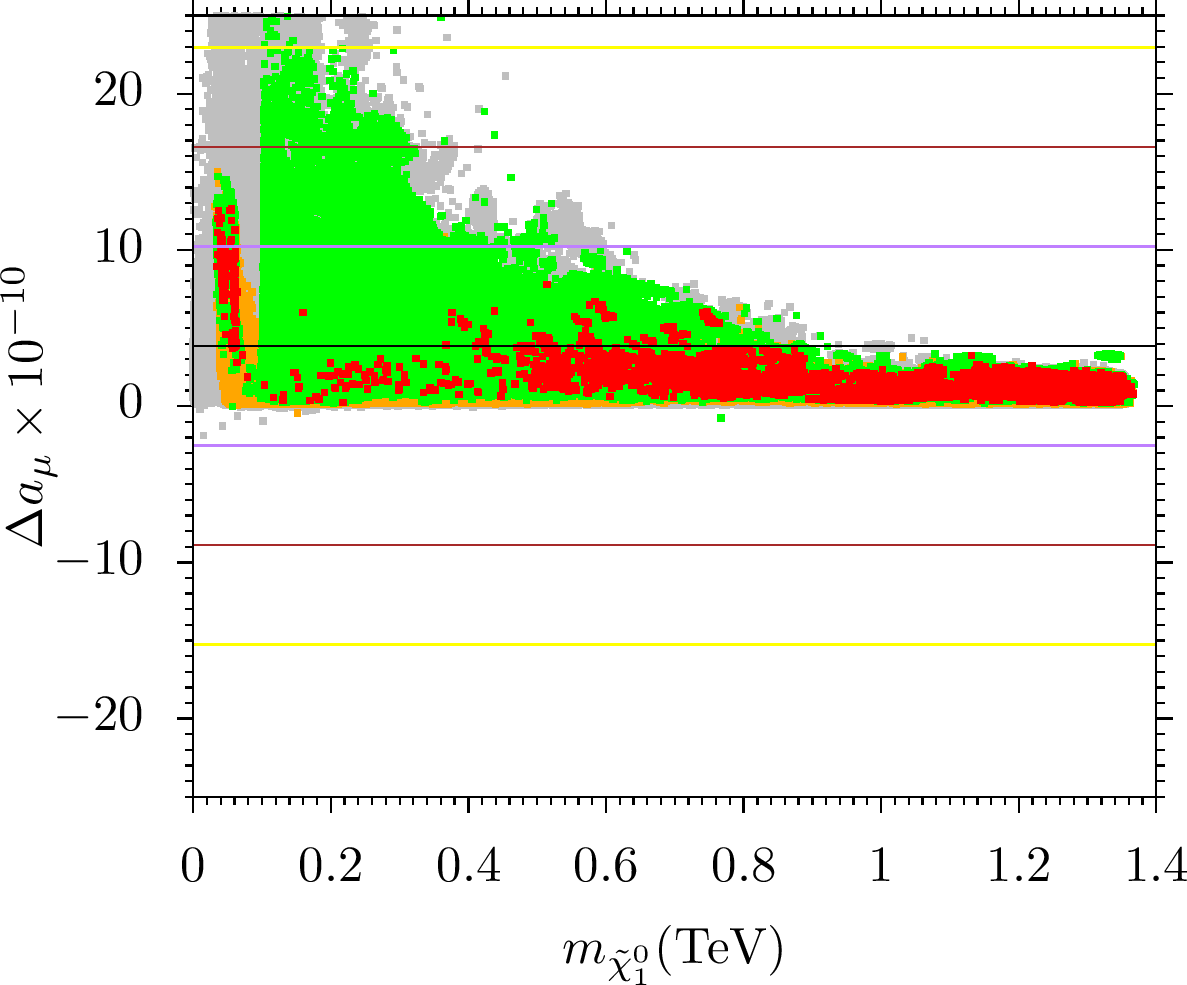}
	\centering \includegraphics[width=7.90cm]{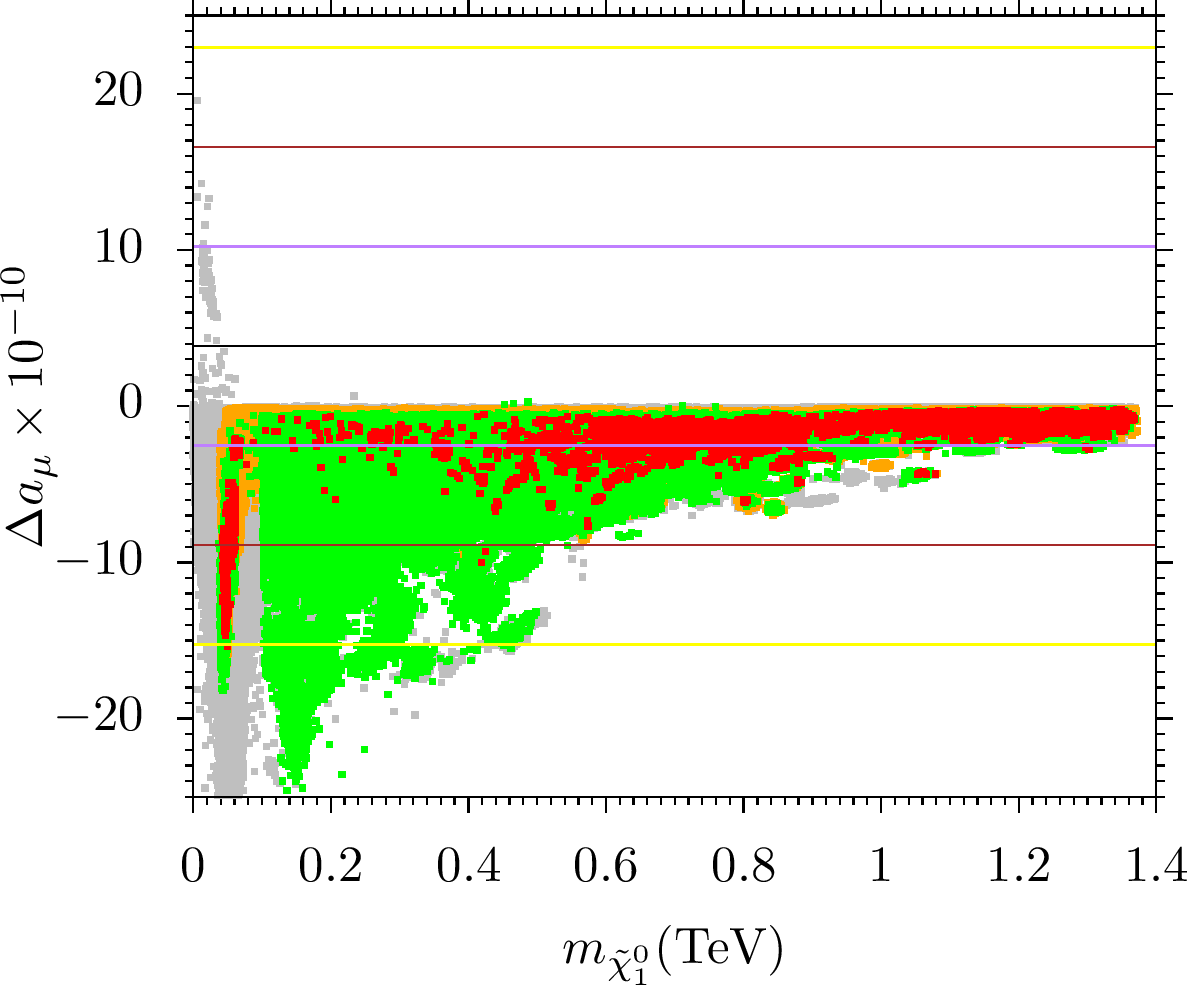}
	\caption{\small  The lightest neutralino DM candidate and the muon anomalous magnetic moment, $a_{\mu}$, is illustrated. Grey points satisfy the REWSB, resulting in LSP neutralino.  Orange points satisfy the LEP mass bounds, B-physics bounds, Higgs bound including S-particles LHC constraints, and satisfy the oversaturated relic density bound. Green points are the subsets of orange points and satisfy underrsaturated relic density bounds. Red points are the subsets of green points that satisfy  the Planck relic density bound. The black curve denotes the central value of $\Delta a_{\mu}$, while the purple, brown, and yellow curves correspond to the $1\sigma$, $2\sigma$, and $3\sigma$ confidence intervals, respectively, around this central value.}
	\label{5}
\end{figure}

\begin{table}[h!]
	\centering
	\scalebox{0.7}{
		\begin{tabular}{|l|cccccc|}
			\hline
			& Point 1 & Point 2 &Point 3& Point 4 &Point 5& Point 6 \\
			\hline
			$m_{0}^{U}$          &   6588      & 112.1&82.2 & 867.5&4212&103.4  \\
			$M_{1},M_{2},M_{3} $         & 2009,3095,380      & 2498,2883,1920.5&2296,2387,2159.5&2616,3039,1981.5&1939,1044,3281.5&2749,3074,2261.5 \\
			$m_{E^c},m_{L}$      &   811,827.8   &1033,692& 988.2,641.9& 973.4,934.9&1173,1146&1142,629.9  \\
			$m_{H_{u}},m_{H_{d}}$           &    1021,4285     & 113.7,2829& 111.9,3448&2513,4430&3013,865.3&126,3321 \\
			$m_{Q},m_{D^{c}},,m_{U^{c}}$    & 6023.1,8505.1,8479.1 & 433.96,144.7,0&410.35,106.10&886,1119.9,789&3874.7,5437.7,5352.7&475.7,133.5,0 \\
			$A_{t}=A_{b},A_{\tau}$            &  15240,2946     & -6218,-1907&-6772,-2076&-5269,5988& -9039,-1298&-7517,-1846 \\
			$\tan\beta$                      & 37.7 & 36.5 & 31.9 & 15&30.3&34.6 \\
   $\mu$                      & -6789.4 & -3146.4 & -3767.2 & -2521&-5866.8&-3788.3 \\
			\hline
			$m_h$            &  125.4    & 124.7&125.7&125.5&125.5&124.4    \\
			$m_H$            &  5474    & 2261&4218 & 5317&4510&2706  \\
			$m_{A} $         &  5439     &  \textbf{2247}& 4191& 5282&4480&2689   \\
			$m_{H^{\pm}}$    &  5475   & 2264 & 4219 & 5318&4511&2708    
 \\
			\hline
			$m_{\tilde{\chi}^0_{1,2}}$
			& \textbf{902},2582 & \textbf{1122},2392& \textbf{1024},1978& \textbf{1171},2465&\textbf{859},866&\textbf{1240},2558 \\
			$m_{\tilde{\chi}^0_{3,4}}$
			& 6694,6694 &3098,3101& 3740,3741& 2525,2572&5775,5776&3731,3733 \\
			$m_{\tilde{\chi}^{\pm}_{1,2}}$
			&2586,6662  & 2394,4100&1985,3743 &2460,2562&\textbf{866},5739&2562,3731  \\
			\hline
			$m_{\tilde{g}}$  & \textbf{988}    &4025& 4471&4147&6846&4686 \\
			\hline $m_{ \tilde{u}_{L,R}}$
			& 6357,8291  & 3929,3494& 4153,3856& 4110,3733&6915,7782&4486,4057  \\
			$m_{\tilde{t}_{1,2}}$
			& 3073,5545  & 1540,2415&1741,2842&\textbf{1224},3262&5216,5617&1861,2627 \\
			\hline $m_{ \tilde{d}_{L,R}}$
			& 6358,8662 & 3930,3500&4152,3898&4111,3761&6916,7926&4487,4068 \\
			$m_{\tilde{b}_{1,2}}$
			& 3112,7114 & 1177,2422&2763,2909&3309,3568&5301,7205&\textbf{1321},2642 \\
			\hline
			$m_{\tilde{\nu}_{1}}$
			& 1237      & 1967&1614&2114&1001&2054 \\
			$m_{\tilde{\nu}_{3}}$
			& 1102      & 1881&1439&2037&1010&1971  \\
			\hline
			$m_{ \tilde{e}_{L,R}}$
			& 1149,2639   & 1969,1479&1620,1469&2117,1542&983,1817&2056,1642 \\
			$m_{\tilde{\tau}_{1,2}}$
			& 1028,2507    & 1226,1885&\textbf{1048},1458&1330,2043&957,1782&1421,1976 \\
			\hline
			$\Delta a_{\mu}$
			& -1.08$\times 10^{-10}$ & -9.9$\times 10^{-11}$ &-1.11$\times 10^{-10}$&-3.2$\times 10^{-11}$&-2$\times 10^{-10}$&-8.3$\times 10^{-11}$    
			 \\
			\hline
			$\sigma_{SI}(pb)$
			& 5.5$\times 10^{-18}$ & 5.1$\times 10^{-12}$&1.6$\times 10^{-13}$&2$\times 10^{-12}$&2$\times 10^{-13}$&8.8$\times 10^{-13}$  \\
   $\sigma_{SD}(pb)$
			& 1.6$\times 10^{-11}$ & 2.9$\times 10^{-10}$ &3.3$\times 10^{-11}$&2.1$\times 10^{-9}$&6.9$\times 10^{-10}$&9.3$\times 10^{-11}$ 
			 \\
			$\Omega_{CDM}h^2$
			& 0.121      & 0.124&0.121&0.125&116&120\\
			\hline
			\hline
		\end{tabular}
  }
 \caption{The input model parameters and the resulting supersymmetric particle (sparticle) mass spectrum are summarized. All masses are provided in units of GeV, corresponding to the scenario with a negative Higgsino mass parameter (\(\mu < 0\)).}
		\label{table1}
\end{table}


\begin{table}[h!]
	\centering
	\scalebox{0.75}{
		\begin{tabular}{|l|ccccc|}
			\hline
			& Point 1 & Point 2 &Point 3& Point 4 &Point 5  \\
			\hline
			$m_{0}^{U}$          &   6710    & 4837&3664 & 5414&2648 \\
			$M_{1},M_{2},M_{3} $         &  1749,2722,289.5      & 2963,2391,3821&2175,2900,1087.5&2555,2232,3039.5&2707,1463,4569\\
			$m_{E^c},m_{L}$      &   224,1107   &730.9,830.5& 186.2,546.2& 664.2,954.2&328.9,1044 \\
			$m_{H_{u}},m_{H_{d}}$           &   260,1593    &  4899,1378&3185,3131&4321,1427&670.1,295.6\\
			$m_{Q},m_{D^{c}},,m_{U^{c}}$     & 6126,8662.6,8660.6 & 4425.6,6244.5,6216&3345.6,4730.2,4727.8&4949.7,6989.4,6968.4&2358.2,3328.2,3314.6 \\
			$A_{t}=A_{b},A_{\tau}$            &   -11770,2881     &-7490,1703&15170,2973&-8342,1580&-10640,-2859 \\
			$\tan\beta$                      & 19&46.7&12.8&56.2\\
   $\mu$                      & 7037.7&5298&4701.1& 5567.4&7299.7\\
			\hline
			$m_h$            &  126.5&125.5&125.5&125.5&125.4   \\
			$m_H$            &  6748&3042&5741&2201&7374 \\
			$m_{A} $         & 6704&3022&5704&\textbf{2186}&7326  \\
			$m_{H^{\pm}}$    &  6748&3043&5742&2203&7374   \\
			\hline
			$m_{\tilde{\chi}^0_{1,2}}$
			&\textbf{805},2326&\textbf{1334},1993&\textbf{965},2390&\textbf{1152},1872&\textbf{1200},1207\\
			$m_{\tilde{\chi}^0_{3,4}}$
			& 7003,7003&5300,5301&4685,5686&5547,5547&7284,7284 \\
			$m_{\tilde{\chi}^{\pm}_{1,2}}$
			&2329,6984&1997,5297&2411,4714&1876,5538&\textbf{1201},7299 \\
			\hline
			$m_{\tilde{g}}$  & \textbf{888}&7863&2467&6446&9137 \\
			\hline $m_{ \tilde{u}_{L,R}}$
			& 6373,8467&8004,8991&4260,5025&7351,8648&8131,8454  \\
			$m_{\tilde{t}_{1,2}}$
			& 4389,5912&6104,6677&\textbf{1007},2389&5099,6342&5812,6943 \\
			\hline $m_{ \tilde{d}_{L,R}}$
			& 6374,8802&8004,9089&4261,5175&7352,8811&8131,8504\\
			$m_{\tilde{b}_{1,2}}$
			& 4465,8500&6186,8145&2340,5003&5190,7529&6940,8398 \\
			\hline
			$m_{\tilde{\nu}_{1}}$
			& 1130&1486&1699&1262&1227  \\
			$m_{\tilde{\nu}_{3}}$
			& 1065&1496&1673&1497&1213 \\
			\hline
			$m_{ \tilde{e}_{L,R}}$
			& 1049,2473&1465,1833&1701,1511&1201,2005&1288,1405 \\
			$m_{\tilde{\tau}_{1,2}}$
			& 1037,2414&\textbf{1413},1798&1449,1678&1400,2184&1222,1413\\
			\hline
			$\Delta a_{\mu}$
			& 7$\times 10^{-11}$ & 1.5$\times 10^{-10}$ &4$\times 10^{-11}$&2.2$\times 10^{-10}$&5.6$\times 10^{-11}$  
			\\
			\hline
			$\sigma_{SI}(pb)$
			& 3.3$\times 10^{-14}$ & 2.2$\times 10^{-13}$&5.2$\times 10^{-13}$&1.8$\times 10^{-13}$&1.1$\times 10^{-12}$  \\
   $\sigma_{SD}(pb)$
			& 1$\times 10^{-11}$ & 9.9$\times 10^{-11}$ &2.6$\times 10^{-11}$&6.9$\times 10^{-11}$&5$\times 10^{-10}$ 
			\\
			$\Omega_{CDM}h^2$
			& 0.117&0.125&0.115&0.125&0.125\\
			\hline
			\hline
		\end{tabular}
  }
 \caption{The input model parameters and the resulting supersymmetric particle (sparticle) mass spectrum are summarized. All masses are provided in units of GeV, corresponding to the scenario with a positive Higgsino mass parameter (\(\mu > 0\)).}
		\label{table2}
\end{table}


\end{widetext}

The upper panels of Fig.~\ref{1} display the parameter space in the $m_{\tilde{t}_1} - m_{\tilde{\chi}_1^0}$ plane, separated into two scenarios: $\mu > 0$ (left) and $\mu < 0$ (right). The gray points satisfy REWSB and assume a neutralino LSP. Orange points satisfy all experimental bounds, including LEP constraints, rare $B$-meson decays, Higgs mass measurements, and LHC sparticle searches, while also being consistent with relic density exceeding the Planck upper bound. A subset of these, marked in green, correspond to relic density values below the Planck limit. Red points represent the most viable solutions consistent with the observed relic density range within $5\sigma$ uncertainty. Black solid curves indicate regions where neutralino-stop coannihilation or resonance effects are expected to dominate. Notably, red points aligned along these curves indicate viable next-to-lightest supersymmetric particle (NLSP) stop scenarios. These solutions occur for $m_{\tilde{t}_1}$ values between approximately $0.15$~TeV and $1.3$~TeV. The apparent sparsity of red, green, and orange points in this region is due to the limited scan density rather than a true exclusion; denser scans would likely populate these regions further. The middle panel presents the mass gap $\Delta m_{\tilde{t}_1,\tilde{\chi}_1^0}$ versus $m_{\tilde{\chi}_1^0}$. For red points, the stop-neutralino mass difference can reach up to 50~GeV. Despite this, the ratio $\Delta m / m_{\tilde{\chi}_1^0}$ remains below $10\%$, consistent with coannihilation requirements. Large mass differences allow kinematically favorable decays such as $\tilde{t}_1 \rightarrow t\, \tilde{\chi}_1^0$, or three- and four-body decays like $\tilde{t}_1 \rightarrow W + b + \tilde{\chi}_1^0$ and $\tilde{t}_1 \rightarrow f + f' + b + \tilde{\chi}_1^0$. In the case of small mass gaps, these decay channels are closed, and the loop-suppressed decay $\tilde{t}_1 \rightarrow c\, \tilde{\chi}_1^0$ becomes dominant~\cite{Hikasa:1987db,Muhlleitner:2011ww}. For recent LHC searches see ~\cite{ATLAS:2017drc,ATLAS:2017eoo,ATLAS:2017www,ATLAS:2017bfj,ATLAS:2019zrq,ATLAS:2020dsf,ATLAS:2020xzu,ATLAS:2021kxv,ATLAS:2021hza}. In all of
these studies, the maximum stop mass considered is 1.2 TeV, which is consistent with the maximum stop mass allowed by all constraints (red points) in our findings, indicating regions potentially accessible in future LHC runs. Even scenarios with small mass splittings where $\tilde{t}_1 \rightarrow t\, \tilde{\chi}_1^0$ the dominant remains viable above 550~GeV, beyond current experimental constraints~\cite{ATLAS:2021kxv}. The lower panels of Fig.~\ref{1} show $m_{\tilde{b}_1}$ as a function of $m_{\tilde{\chi}_1^0}$ for $\mu > 0$ and $\mu < 0$. The same color scheme applies. Blue points additionally satisfy flavor constraints from rare $B$-meson processes. The diagonal line corresponds to mass-degenerate regions indicative of sbottom-neutralino coannihilation. In our current analysis, no sbottom-coannihilation is observed for $\mu > 0$. However, in the $\mu < 0$ case, viable red points near 1--1.3~TeV hint at possible sbottom coannihilation. In our present study, somehow we do not have a large density of green and orange points in these
channels, so in the results, no red points. This is an artifact of scanning. Had we done some more
focused scans, we would have populated this region of parameter space with more points, so we
would get NLSP sbottom solutions too. For $\Delta m_{\tilde{b}_1,\tilde{\chi}_1^0} > m_b$, dominant decays include $pp \rightarrow \tilde{b}_1 \tilde{b}_1^* X \rightarrow b \bar{b} + \not\!\!E_T$,
with $\tilde{b}_1 \rightarrow b\, \tilde{\chi}_1^0$ being the primary decay channel. Additional signatures may arise from same-sign sbottom pair production: $\tilde{b}_1 \tilde{b}_1$ or $\tilde{b}_1^* \tilde{b}_1^*$. Recent ATLAS results have excluded sbottom masses up to 1.5~TeV for decay chains involving heavier neutralinos or charginos~\cite{ATLAS:2019gdh,ATLAS:2021pzz,ATLAS:2019fag}. For compressed spectra with $\Delta m \sim 10$~GeV, searches employing secondary vertex tagging techniques exclude sbottoms up to 660~GeV~\cite{ATLAS:2021yij}. Monojet searches have ruled out masses up to 600~GeV. Importantly, for $m_{\tilde{b}_1} > 800$~GeV and small mass gaps, there are no stringent bounds~\cite{ATLAS:2021yij}, leaving our high-mass sbottom solutions unconstrained and promising for discovery in Run-3 and future colliders.

Figure~\ref{2} presents our findings in the $m_{\tilde{\tau}_1}$--$m_{\tilde{\chi}_1^0}$ (top panel) and $m_{\tilde{\tau}_1}$--$|\Delta m_{\tilde{\tau}_1,\tilde{\chi}_1^0}|$ (middle panel) parameter planes. The color coding is consistent with that used in Figure~\ref{1}. The top panel illustrates the stau-neutralino coannihilation region, while the middle panel highlights the corresponding mass splitting between the lightest stau ($\tilde{\tau}_1$) and the lightest neutralino ($\tilde{\chi}_1^0$). From our parameter scan, we observe that viable coannihilation solutions—characterized by near mass degeneracy between $\tilde{\tau}_1$ and $\tilde{\chi}_1^0$—are located within the range $0.15~\text{TeV} \lesssim m_{\tilde{\tau}_1} \lesssim 1.4~\text{TeV}$. These findings are consistent with the experimental results reported by the CMS Collaboration based on $137~\mathrm{fb}^{-1}$ data collected at $\sqrt{s} = 13$ TeV~\cite{CMS:2022rqk}. We anticipate that future data from LHC Run 3 and high-luminosity phases will be capable of probing a significant portion of this parameter space. In addition to coannihilation scenarios, our scan also reveals the presence of Higgs funnel solutions. In these regions, the relic abundance of neutralino dark matter is effectively reduced through $s$-channel resonant annihilation via CP-even or CP-odd Higgs bosons, specifically $h$, $H$, and $A$, with the condition $m_A \simeq 2 m_{\tilde{\chi}_1^0}$. We also note that in this scenario $m_A \simeq m_H$. Our results indicate that such funnel solutions are present for $m_A$ values ranging from approximately $1.2~\text{TeV}$ to $2.9~\text{TeV}$. Recent constraints from the CMS experiment~\cite{CMS:2022goy} demonstrate that for the decay mode $A \to \tau^+ \tau^-$, pseudoscalar Higgs masses below $1.7~\text{TeV}$ are excluded for $\tan\beta \lesssim 30$. Furthermore, projections from Refs.~\cite{Baer:2022qqr,Baer:2022smj} indicate that for $\tan\beta \lesssim 10$, values of $m_A \sim 1.0$, $1.1$, and $1.4~\text{TeV}$ may be testable at Run 2, Run 3, and the HL-LHC, respectively. Our parameter space includes scenarios that lie beyond these current exclusion limits, thereby motivating continued experimental exploration in this regime.

In addition to the previously discussed annihilation channels, our parameter space scan reveals viable chargino-neutralino coannihilation regions. These are illustrated in Figure~\ref{3}, where we present the distributions in the $m_{\tilde{\chi}_1^\pm}$--$m_{\tilde{\chi}_1^0}$ (top panel) and $m_{\tilde{\chi}_1^\pm}$--$|\Delta m_{\tilde{\chi}_1^\pm,\tilde{\chi}_1^0}|$ (middle panel) planes. The red data points correspond to scenarios where the lightest chargino ($\tilde{\chi}_1^\pm$) is nearly degenerate in mass with the lightest neutralino ($\tilde{\chi}_1^0$), lying within the mass interval $0.3~\text{TeV} \lesssim m_{\tilde{\chi}_1^\pm} \lesssim 1.5~\text{TeV}$. Experimental bounds from direct searches for electroweakinos are provided in Ref.~\cite{ATLAS:2021ilc}, which investigates final states arising from slepton- and $W/Z/h$-mediated decays of chargino and neutralino pairs. According to these constraints, degenerate scenarios with $m_{\tilde{\chi}_1^\pm} \gtrsim 300$~GeV are not excluded at the 95\% confidence level, thereby validating the heavier solutions in our scan. Furthermore, in regions of the parameter space where sleptons are heavier than charginos, decay channels mediated by virtual sleptons are kinematically suppressed, favoring decays through Standard Model bosons. This supports the viability of our heavy chargino-neutralino coannihilation solutions in models where the slepton sector is decoupled. The bottom panel of Figure~\ref{3} illustrates the light Higgs boson mass ($m_h$) and the higgsino mass parameter ($\mu$). Our viable solutions cluster around the observed Higgs mass $m_h \approx 125$~GeV, consistent with current experimental measurements. Notably, the parameter $\mu$ spans a broad range, reaching up to approximately $10$~TeV, with both moderate and large $\mu$ solutions contributing to the viable regions. The presence of such solutions offers promising targets for future high-energy collider experiments.

The top panel of Figure~\ref{4} illustrates the correlation between the mass of the lightest neutralino ($m_{\tilde{\chi}_1^0}$) and that of the NLSP gluino ($m_{\tilde{g}}$). The color scheme follows that of Figure~\ref{3}, with the exception that here we omit any imposed lower bounds on the gluino mass, as discussed in Section~\ref{sec:scan}. The diagonal clustering of red points in the mass range $0.2~\text{TeV} \lesssim m_{\tilde{g}} \lesssim 1.2~\text{TeV}$ reflects a highly compressed spectrum, wherein the gluino is nearly degenerate with the neutralino. In such scenarios, the decay mode $\tilde{g} \rightarrow b\bar{b}\tilde{\chi}_1^0$ tends to dominate. Although other decay modes are kinematically viable, they suffer from overwhelming Standard Model background contamination, particularly at low jet transverse momentum ($p_T$). The advantage of the $b$-quark final state lies in its experimental signature: secondary vertices and displaced tracks from $b$-hadron decays can be reconstructed even when the energy is modest, enabling identification of soft $b$-jets. In contrast, decay channels involving lighter quarks yield jets that often fail to pass track or vertex reconstruction thresholds, reducing their utility in compressed searches. Experimental analyses from the ATLAS Collaboration~\cite{ATLAS:2021ilc,ATLAS:2022ihe} have placed stringent exclusion limits on this compressed topology. Specifically, scenarios in which $m_{\tilde{g}} \approx m_{\tilde{\chi}_1^0}$ are excluded for gluino masses up to approximately 1.2~TeV. Consequently, a significant portion of the red solutions presented in this region are already ruled out by LHC Run 2 data, emphasizing the experimental pressure on such compressed supersymmetric spectra. Finally, in Figure~\ref{4}, we present the predicted neutralino-nucleon scattering cross sections from our parameter scan, where the middle panel corresponds to the SI cross section $\sigma_{\text{SI}}$ and the bottom panel to the SD cross section $\sigma_{\text{SD}}$, both as functions of the lightest neutralino mass $m_{\tilde{\chi}_1^0}$. Experimental exclusion bounds are overlaid for comparison: the solid black line denotes current XENONnT limits~\cite{XENON:2023cxc}, while the solid purple and blue lines represent the 2022 and projected 1000-day sensitivity limits from the LUX-ZEPLIN (LZ) experiment, respectively~\cite{LZ:2018qzl, LZ:2022lsv}. As evident in the $m_{\tilde{\chi}_1^0}$--$\sigma_{\text{SI}}$ plane, the majority of solutions (marked in red) lie safely below the exclusion thresholds set by XENONnT and the LZ 2022 results. Furthermore, a large fraction of these points fall within the future sensitivity range of the LZ 1000-day dataset, highlighting their experimental relevance. Apart from a few isolated points around $m_{\tilde{\chi}_1^0} \sim 0.8~\text{TeV}$, nearly all red points evade current exclusion limits. Our scan includes resonance solutions where the LSP mass is close to half the mass of the SM Higgs boson ($m_h/2$) or the $Z$ boson ($m_Z/2$), corresponding to the so-called Higgs and $Z$-funnel (pole) regions. In such cases, the LSP may contribute to the invisible decay widths of these bosons. For $\mu > 0$, we observe that the current LZ constraints already rule out $Z$-pole solutions and exclude portions of the Higgs-pole region (top left panel). In contrast, for $\mu < 0$, both the $Z$- and $H$-pole scenarios, they remain largely viable (top right panel), indicating that the negative $\mu$ region retains significant phenomenological interest and merits focused study in light of upcoming experimental results. The $m_{\tilde{\chi}_1^0}$--$\sigma_{\text{SD}}$ plot (bottom panel) reveals that all viable solutions remain consistent with existing limits and projected future sensitivities for SD scattering. These findings collectively underscore the complementarity of SI and SD detection modes in constraining supersymmetric DM scenarios. 

There has been a persistent discrepancy between the Standard Model (SM) prediction and the experimental measurement of the muon's anomalous magnetic moment, defined as \( a_\mu \equiv (g-2)_\mu/2 \). The SM prediction reported in the 2020 White Paper (WP20) \cite{Aoyama:2020ynm} was
based on data-driven evaluations of the leading-order hadronic vacuum polarization (LO HVP) using \( e^+e^- \) scattering data. 
In 2025, a significant update to the SM prediction was released, adopting lattice QCD results for the LO HVP contribution \cite{Aliberti:2025beg}, this update leads to an upward shift in the SM central value.
The most recent combined experimental measurement from E989 and E821 is given by \cite{Muong-2:2023cdq}.
This yields an updated difference between experiment and theory \cite{Aliberti:2025beg},
corresponding to a statistical deviation of only \( 0.6 \sigma \), compared to the earlier \cite{Muong-2:2021ojo}.
The significant reduction in the discrepancy arises primarily from the adoption of the lattice QCD-based LO HVP contribution, which replaces the tension-prone \( e^+e^- \) data-driven input. As a result, the experimental measurement and SM prediction are now in good agreement, and the previously strong indication of new physics has substantially weakened. Within the framework of SUSY GUTs, the sign of the Higgsino mass parameter \( \mu \) plays a critical role in determining the supersymmetric contribution to \( a_\mu \). Figure~\ref{5} presents the results from a comprehensive scan over the GmSUGRA parameter space. Our analysis reveals that in the \( \mu < 0 \) regime, the supersymmetric contribution to \( (g-2)_\mu \) remains consistent with the experimental measurement within \( 1\sigma \) to \( 2\sigma \) uncertainty. Notably, viable solutions are found with neutralino masses extending up to 1.3~TeV that contribute within the \( 1\sigma \) range. Additionally, the Higgs/\( Z \)-poles in this scenario yield contributions consistent within \( 2\sigma \) while satisfying the Planck 2018 relic density constraints, as represented by the red points. On the other hand, in the \( \mu > 0 \) scenario, although the Higgs and \( Z \)-funnel regions also produce contributions within \( 1\sigma \), these parameter spaces are already ruled out by the latest LUX-ZEPLIN (LZ) direct detection limits. These results highlight the phenomenological viability of the \( \mu < 0 \) scenario under current experimental bounds. These findings underscore the importance of further phenomenological studies of SUSY GUT models under the \( \mu < 0 \) regime, as it remains consistent with both the updated \( (g-2)_\mu \) measurements in the SM and existing experimental bounds.

We conclude by presenting two benchmark tables. Table~\ref{table1} summarizes representative scenarios for the case with a negative Higgsino mass parameter (\(\mu < 0\)), while Table~\ref{table2} corresponds to the case with \(\mu > 0\). In Table~\ref{table1}, \textbf{Point 1} illustrates a scenario with a gluino as the NLSP, which has already been ruled out by current LHC SUSY searches. \textbf{Point 2} represents a Higgs funnel (resonance) solution, where the pseudoscalar Higgs mass \(m_A = 2.247~\text{TeV}\) and the heavy scalar Higgs mass \(m_H = 2.264~\text{TeV}\) are nearly degenerate. The LSP in this case is a bino-like neutralino with mass approximately \(0.902~\text{TeV}\), making it a valid \(A/H\)-resonance solution. \textbf{Point 3} corresponds to an NLSP stau scenario in which the stau mass is approximately \(1.048~\text{TeV}\), and the LSP, a bino-like neutralino, has a mass around \(1.024~\text{TeV}\). \textbf{Point 4} exemplifies stop-neutralino co-annihilation, with the NLSP stop mass around \(1.224~\text{TeV}\) and the LSP neutralino (bino-like) mass close to \(1.171~\text{TeV}\). \textbf{Point 5} features chargino-neutralino co-annihilation, where the LSP neutralino, dominantly bino with some wino admixture, has a mass of approximately \(0.859~\text{TeV}\), while the lightest chargino mass is around \(0.866~\text{TeV}\). Finally, \textbf{Point 6} showcases sbottom-neutralino co-annihilation, with the NLSP sbottom at \(1.321~\text{TeV}\) and the bino-like LSP neutralino at \(1.240~\text{TeV}\). These solutions are consistent with relic density constraints and remain viable under current collider and DM search limits.

Table~\ref{table2} presents representative benchmark points corresponding to the \(\mu > 0\) scenario. \textbf{Point 1} features a case with the gluino as the NLSP, which is already excluded by current LHC SUSY search limits. \textbf{Point 2} represents an NLSP stau scenario, where the stau has a mass of approximately \(1.413~\text{TeV}\), and the LSP, a bino-like neutralino, has a mass of about \(1.133~\text{TeV}\). \textbf{Point 3} illustrates stop-neutralino co-annihilation, with the NLSP stop mass around \(1.007~\text{TeV}\) and the LSP neutralino (bino-dominated) at approximately \(0.965~\text{TeV}\). \textbf{Point 4} corresponds to a Higgs funnel (resonance) scenario, characterized by nearly degenerate heavy Higgs bosons with \(m_A = 2.186~\text{TeV}\) and \(m_H = 2.202~\text{TeV}\), while the LSP neutralino, predominantly bino, has a mass close to \(1.152~\text{TeV}\). This near-degeneracy allows us to interpret it as an \(A/H\)-resonance solution. \textbf{Point 5} depicts a chargino-neutralino co-annihilation scenario, where the lightest chargino has a mass of \(1.202~\text{TeV}\), and the LSP neutralino, which is primarily bino with admixtures of wino and higgsino, has a mass around \(1.200~\text{TeV}\). All scenarios yield viable DM relic densities and are consistent with current collider and DD constraints.

All benchmark points comply with existing bounds from LHC SUSY searches and DD DM experiments, except for Point 1 in both cases. The remaining scenarios could be probed in the LHC Run-3, future SUSY searches, and  DD DM experiments.

\section{Conclusion:}
In this study, we have performed a comprehensive investigation of the parameter space of the GmSUGRA model within the framework of the MSSM, considering positive as well as negative signs of the higgsino mass parameter \(\mu\), as the anomalous magnetic moment of the muon may now be consistent with the SM prediction. Our findings indicate that, in the \( \mu < 0 \) scenario, the supersymmetric contributions to the muon's anomalous magnetic moment remain within a \( 2\sigma \) deviation from the central experimental value, taking into account the current Standard Model prediction. Moreover, our finding reveals that the \(\mu < 0\) configuration accommodates a wider and more flexible set of phenomenologically viable solutions, including sbottom-neutralino coannihilation channels that are absent for \(\mu > 0\). The DM production mechanisms explored encompass coannihilation with gluinos, staus, stops, sbottoms, and charginos, as well as annihilation through the \(A\)-, \(H\)-, and \(Z\)-boson resonance channels. While the \(\mu > 0\) scenario is significantly constrained by current limits from LUX-ZEPLIN (LZ) DD data in the H/Z pole regions while the \(\mu < 0\) region remains largely viable and phenomenologically appealing. Notably, several benchmark points identified in our scan are within the projected discovery reach of LHC Run-3 and the upcoming 1000-day exposure of the LZ experiment. These findings strongly motivate further experimental efforts focused on probing the electroweak sector of SUSY, especially in the context of the negative \(\mu\) scenario.

\textbf{Acknowledgments:--} TL is supported in part by the National Key Research and Development Program of China Grant No. 2020YFC2201504, by the Projects No. 11875062, No. 11947302, No. 12047503, and No. 12275333 supported by the National Natural Science Foundation of China, by the Key Research Program of the Chinese Academy of Sciences, Grant NO. XDPB15, by the Scientific Instrument Developing Project of the Chinese Academy of Sciences, Grant No. YJKYYQ20190049, and by the International Partnership Program of Chinese Academy of Sciences for Grand Challenges, Grant No. 112311KYSB20210012.



\begin{thebibliography}{}



\bibitem{gaugeunification} 

S.~Dimopoulos, S.~Raby and F.~Wilczek,
Phys. Rev. D \textbf{24}, 1681-1683 (1981)
doi:10.1103/PhysRevD.24.1681;
U.~Amaldi, W.~de Boer and H.~Furstenau,
Phys. Lett. B \textbf{260}, 447-455 (1991)
doi:10.1016/0370-2693(91)91641-8;
J.~R.~Ellis, S.~Kelley and D.~V.~Nanopoulos,
Phys. Lett. B \textbf{260}, 131-137 (1991)
doi:10.1016/0370-2693(91)90980-5;
P.~Langacker,
J. Phys. G \textbf{29}, 35-48 (2003)
doi:10.1088/0954-3899/29/1/305
[arXiv:hep-ph/0102085 [hep-ph]].

\bibitem{Georgi:1974sy}
H.~Georgi and S.~L.~Glashow,
Phys. Rev. Lett. \textbf{28}, 1494 (1972)
doi:10.1103/PhysRevLett.28.1494

\bibitem{Pati:1974yy}
J.~C.~Pati and A.~Salam,
Phys. Rev. D \textbf{10}, 275-289 (1974)
[erratum: Phys. Rev. D \textbf{11}, 703-703 (1975)]
doi:10.1103/PhysRevD.10.275

\bibitem{Mohapatra:1974hk}
R.~N.~Mohapatra and J.~C.~Pati,
Phys. Rev. D \textbf{11}, 2558 (1975)
doi:10.1103/PhysRevD.11.2558

\bibitem{Fritzsch:1974nn}
H.~Fritzsch and P.~Minkowski,
Annals Phys. \textbf{93}, 193-266 (1975)
doi:10.1016/0003-4916(75)90211-0

\bibitem{Georgi:1974my}
H.~Georgi,
AIP Conf. Proc. \textbf{23}, 575-582 (1975)
doi:10.1063/1.2947450

\bibitem{neutralinodarkmatter}
H.~Goldberg,
Phys. Rev. Lett. \textbf{50}, 1419 (1983)
[erratum: Phys. Rev. Lett. \textbf{103}, 099905 (2009)]
doi:10.1103/PhysRevLett.50.1419;
J.~R.~Ellis, J.~S.~Hagelin, D.~V.~Nanopoulos, K.~A.~Olive and M.~Srednicki,
Nucl. Phys. B \textbf{238}, 453-476 (1984)
doi:10.1016/0550-3213(84)90461-9.

\bibitem{darkmatterreviews} For reviews, see
G.~Jungman, M.~Kamionkowski and K.~Griest,
Phys. Rept. \textbf{267}, 195-373 (1996)
doi:10.1016/0370-1573(95)00058-5
[arXiv:hep-ph/9506380 [hep-ph]];
K.~A.~Olive,
[arXiv:astro-ph/0301505 [astro-ph]];
J.~L.~Feng,
eConf \textbf{C0307282}, L11 (2003)
[arXiv:hep-ph/0405215 [hep-ph]];
M.~Drees,
AIP Conf. Proc. \textbf{805}, no.1, 48-54 (2005)
doi:10.1063/1.2149675
[arXiv:hep-ph/0509105 [hep-ph]];
J.~L.~Feng,
Ann. Rev. Astron. Astrophys. \textbf{48}, 495-545 (2010)
doi:10.1146/annurev-astro-082708-101659
[arXiv:1003.0904 [astro-ph.CO]].


\bibitem{Slavich:2020zjv}
P.~Slavich, S.~Heinemeyer, E.~Bagnaschi, H.~Bahl, M.~Goodsell, H.~E.~Haber, T.~Hahn, R.~Harlander, W.~Hollik and G.~Lee, \textit{et al.}
Eur. Phys. J. C \textbf{81} (2021) no.5, 450
doi:10.1140/epjc/s10052-021-09198-2
[arXiv:2012.15629 [hep-ph]].
\bibitem{ATLAS-SUSY-Search}
ATLAS Collaboration, ATLAS-CONF-2019-040.

\bibitem{Aad:2020sgw}
G.~Aad \textit{et al.} [ATLAS],
Eur. Phys. J. C \textbf{80}, no.8, 737 (2020)
doi:10.1140/epjc/s10052-020-8102-8
[arXiv:2004.14060 [hep-ex]].

\bibitem{Aad:2019pfy}
G.~Aad \textit{et al.} [ATLAS],
JHEP \textbf{12}, 060 (2019)
doi:10.1007/JHEP12(2019)060
[arXiv:1908.03122 [hep-ex]].

\bibitem{CMS-SUSY-Search-I}
CMS Collaboration, CMS PAS SUS-19-005.

\bibitem{CMS-SUSY-Search-II}
CMS Collaboration, CMS PAS SUS-19-006.


\bibitem{Ahmed:2022ude}
W.~Ahmed, I.~Khan, T.~Li, S.~Raza and W.~Zhang,
Phys. Lett. B \textbf{832} (2022), 137216
doi:10.1016/j.physletb.2022.137216
[arXiv:2202.11011 [hep-ph]].

\bibitem{Zhang:2023jcf}
W.~Zhang, W.~Ahmed, I.~Khan, T.~Li and S.~Raza,
Phys. Rev. D \textbf{110}, no.5, 055006 (2024)
doi:10.1103/PhysRevD.110.055006
[arXiv:2304.01082 [hep-ph]].

\bibitem{Khan:2023ryc}
I.~Khan, W.~Ahmed, T.~Li and S.~Raza,
Phys. Rev. D \textbf{109} (2024) no.7, 075051
doi:10.1103/PhysRevD.109.075051
[arXiv:2312.07863 [hep-ph]].



\bibitem{ATLAS:2020zms}
G.~Aad \textit{et al.} [ATLAS],
Phys. Rev. Lett. \textbf{125} (2020) no.5, 051801
doi:10.1103/PhysRevLett.125.051801
[arXiv:2002.12223 [hep-ex]].

\bibitem{CMS:2020bfa}
A.~M.~Sirunyan \textit{et al.} [CMS],
JHEP \textbf{04} (2021), 123
doi:10.1007/JHEP04(2021)123
[arXiv:2012.08600 [hep-ex]].

\bibitem{ATLAS:2021moa}
G.~Aad \textit{et al.} [ATLAS],
Eur. Phys. J. C \textbf{81} (2021) no.12, 1118
doi:10.1140/epjc/s10052-021-09749-7
[arXiv:2106.01676 [hep-ex]].

\bibitem{ATLAS:2021yqv}
G.~Aad \textit{et al.} [ATLAS],
Phys. Rev. D \textbf{104} (2021) no.11, 112010
doi:10.1103/PhysRevD.104.112010
[arXiv:2108.07586 [hep-ex]].

\bibitem{CMS:2022sfi}
A.~Tumasyan \textit{et al.} [CMS],
Phys. Lett. B \textbf{842} (2023), 137460
doi:10.1016/j.physletb.2022.137460
[arXiv:2205.09597 [hep-ex]].

\bibitem{ATLAS:2022yvh}
G.~Aad \textit{et al.} [ATLAS],
JHEP \textbf{08} (2022), 104
doi:10.1007/JHEP08(2022)104
[arXiv:2202.07953 [hep-ex]].


\bibitem{XENON:2018voc}
E.~Aprile \textit{et al.} [XENON],
Phys. Rev. Lett. \textbf{121} (2018) no.11, 111302
doi:10.1103/PhysRevLett.121.111302
[arXiv:1805.12562 [astro-ph.CO]].

\bibitem{XENON:2019rxp}
E.~Aprile \textit{et al.} [XENON],
Phys. Rev. Lett. \textbf{122} (2019) no.14, 141301
doi:10.1103/PhysRevLett.122.141301
[arXiv:1902.03234 [astro-ph.CO]].

\bibitem{PICO:2019vsc}
C.~Amole \textit{et al.} [PICO],
Phys. Rev. D \textbf{100} (2019) no.2, 022001
doi:10.1103/PhysRevD.100.022001
[arXiv:1902.04031 [astro-ph.CO]].

\bibitem{PandaX-4T:2021bab}
Y.~Meng \textit{et al.} [PandaX-4T],
Phys. Rev. Lett. \textbf{127} (2021) no.26, 261802
doi:10.1103/PhysRevLett.127.261802
[arXiv:2107.13438 [hep-ex]].

\bibitem{LZ:2022lsv}
J.~Aalbers \textit{et al.} [LZ],
Phys. Rev. Lett. \textbf{131} (2023) no.4, 041002
doi:10.1103/PhysRevLett.131.041002
[arXiv:2207.03764 [hep-ex]].

\bibitem{LZ:2024zvo}
J.~Aalbers \textit{et al.} [LZ],
[arXiv:2410.17036 [hep-ex]].

\bibitem{PandaX:2022xas}
Z.~Huang \textit{et al.} [PandaX],
Phys. Lett. B \textbf{834} (2022), 137487
doi:10.1016/j.physletb.2022.137487
[arXiv:2208.03626 [hep-ex]].


\bibitem{KhalilS2017}
Khalil, S., $\&$ Moretti, S. (2017). Supersymmetry beyond minimality: from theory to experiment. CRC Press.

\bibitem{Ahmed:2021htr}
W.~Ahmed, I.~Khan, J.~Li, T.~Li, S.~Raza and W.~Zhang,
Phys. Lett. B \textbf{827} (2022), 136879
doi:10.1016/j.physletb.2022.136879
[arXiv:2104.03491 [hep-ph]].

\bibitem{Muong-2:2023cdq}
D.~P.~Aguillard \textit{et al.} [Muon g-2],
Phys. Rev. Lett. \textbf{131}, no.16, 161802 (2023)
doi:10.1103/PhysRevLett.131.161802
[arXiv:2308.06230 [hep-ex]].

\bibitem{Muong-2:2021ojo}
B.~Abi \textit{et al.} [Muon g-2],
Phys. Rev. Lett. \textbf{126}, no.14, 141801 (2021)
doi:10.1103/PhysRevLett.126.141801
[arXiv:2104.03281 [hep-ex]].

\bibitem{Muong-2:2025xyk}
D.~P.~Aguillard \textit{et al.} [Muon g-2],
[arXiv:2506.03069 [hep-ex]].

\bibitem{Aliberti:2025beg}
R.~Aliberti, T.~Aoyama, E.~Balzani, A.~Bashir, G.~Benton, J.~Bijnens, V.~Biloshytskyi, T.~Blum, D.~Boito and M.~Bruno, \textit{et al.}
[arXiv:2505.21476 [hep-ph]].


\bibitem{Barman:2022jdg}
R.~K.~Barman, G.~B\'elanger, B.~Bhattacherjee, R.~M.~Godbole and R.~Sengupta,
Phys. Rev. Lett. \textbf{131}, no.1, 011802 (2023)
doi:10.1103/PhysRevLett.131.011802
[arXiv:2207.06238 [hep-ph]].

\bibitem{Khan:2025azf}
I.~Khan, W.~Ahmed, T.~Li, S.~Raza and A.~Muhammad,
[arXiv:2501.12039 [hep-ph]].

\bibitem{Li:2010xr}
T.~Li and D.~V.~Nanopoulos,
Phys. Lett. B \textbf{692}, 121-125 (2010)
doi:10.1016/j.physletb.2010.07.024
[arXiv:1002.4183 [hep-ph]].

\bibitem{Balazs:2010ha}
C.~Balazs, T.~Li, D.~V.~Nanopoulos and F.~Wang,
JHEP \textbf{09}, 003 (2010)
doi:10.1007/JHEP09(2010)003
[arXiv:1006.5559 [hep-ph]].

\bibitem{Cheng:2012np}
T.~Cheng, J.~Li, T.~Li, D.~V.~Nanopoulos and C.~Tong,
Eur. Phys. J. C \textbf{73}, 2322 (2013)
doi:10.1140/epjc/s10052-013-2322-0
[arXiv:1202.6088 [hep-ph]].

\bibitem{ISAJET}
H.~Baer, F.~E.~Paige, S.~D.~Protopopescu and X.~Tata,
[arXiv:hep-ph/0001086 [hep-ph]].

\bibitem{ATLAS:2014wva}
 [ATLAS, CDF, CMS and D0],
[arXiv:1403.4427 [hep-ex]].

\bibitem{Belanger:2009ti}
  G.~Belanger, F.~Boudjema, A.~Pukhov and R.~K.~Singh,
JHEP \textbf{11}, 026 (2009)
doi:10.1088/1126-6708/2009/11/026
[arXiv:0906.5048 [hep-ph]].
H.~Baer, S.~Kraml, S.~Sekmen and H.~Summy,
JHEP \textbf{03}, 056 (2008)
doi:10.1088/1126-6708/2008/03/056
[arXiv:0801.1831 [hep-ph]].

\bibitem{Patrignani:2016xqp}
R.~L.~Workman \textit{et al.} [Particle Data Group],
PTEP \textbf{2022}, 083C01 (2022)
doi:10.1093/ptep/ptac097

\bibitem{Khachatryan:2016vau} 
  G.~Aad {\it et al.} [ATLAS and CMS Collaborations],
  JHEP {\bf 1608}, 045 (2016)
  doi:10.1007/JHEP08(2016)045
  [arXiv:1606.02266 [hep-ex]].

\bibitem{Allanach:2004rh} 
  B.~C.~Allanach, A.~Djouadi, J.~L.~Kneur, W.~Porod and P.~Slavich,
  JHEP {\bf 0409}, 044 (2004)
doi:10.1088/1126-6708/2004/09/044

\bibitem{Baer:1997jq}
H.~Baer, M.~Brhlik, D.~Castano and X.~Tata,
Phys. Rev. D \textbf{58}, 015007 (1998)
doi:10.1103/PhysRevD.58.015007
[arXiv:hep-ph/9712305 [hep-ph]].

\bibitem{Babu:1999hn}
K.~S.~Babu and C.~F.~Kolda,
Phys. Rev. Lett. \textbf{84}, 228-231 (2000)
doi:10.1103/PhysRevLett.84.228
[arXiv:hep-ph/9909476 [hep-ph]].

\bibitem{Dedes:2001fv}
A.~Dedes, H.~K.~Dreiner and U.~Nierste,
Phys. Rev. Lett. \textbf{87}, 251804 (2001)
doi:10.1103/PhysRevLett.87.251804
[arXiv:hep-ph/0108037 [hep-ph]].

\bibitem{Mizukoshi:2002gs}
J.~K.~Mizukoshi, X.~Tata and Y.~Wang,
Phys. Rev. D \textbf{66}, 115003 (2002)
doi:10.1103/PhysRevD.66.115003
[arXiv:hep-ph/0208078 [hep-ph]].

\bibitem{Aaij:2012nna} 
  R.~Aaij {\it et al.} [LHCb Collaboration],
  Phys.\ Rev.\ Lett.\  {\bf 110}, no. 2, 021801 (2013)
 doi:10.1103/PhysRevLett.110.021801
  [arXiv:1211.2674 [hep-ex]].

\bibitem{Amhis:2012bh} 
  Y.~Amhis \textit{et al.} [HFLAV],
[arXiv:1207.1158 [hep-ex]].

\bibitem{Asner:2010qj} 
  D.~Asner \textit{et al.} [HFLAV],
[arXiv:1010.1589 [hep-ex]].
\bibitem{Akrami:2018vks} 
N.~Aghanim \textit{et al.} [Planck],
Astron. Astrophys. \textbf{641}, A1 (2020)
doi:10.1051/0004-6361/201833880
[arXiv:1807.06205 [astro-ph.CO]].

\bibitem{XENON:2023cxc}
E.~Aprile \textit{et al.} [XENON],
Phys. Rev. Lett. \textbf{131}, no.4, 041003 (2023)
doi:10.1103/PhysRevLett.131.041003
[arXiv:2303.14729 [hep-ex]].

\bibitem{LZ:2018qzl}
D.~S.~Akerib \textit{et al.} [LZ],
Phys. Rev. D \textbf{101}, no.5, 052002 (2020)
doi:10.1103/PhysRevD.101.052002
[arXiv:1802.06039 [astro-ph.IM]].





\bibitem{Hikasa:1987db}
K.~i.~Hikasa and M.~Kobayashi,
Phys. Rev. D \textbf{36}, 724 (1987).

\bibitem{Muhlleitner:2011ww}
M.~Muhlleitner and E.~Popenda,
JHEP \textbf{04}, 095 (2011)
doi:10.1007/JHEP04(2011)095
[arXiv:1102.5712 [hep-ph]].


\bibitem{ATLAS:2017drc}
M.~Aaboud \textit{et al.} [ATLAS],
JHEP \textbf{12}, 085 (2017)
[arXiv:1709.04183 [hep-ex]].

\bibitem{ATLAS:2017eoo}
M.~Aaboud \textit{et al.} [ATLAS],
JHEP \textbf{06}, 108 (2018)
[arXiv:1711.11520 [hep-ex]].


\bibitem{ATLAS:2017www}
M.~Aaboud \textit{et al.} [ATLAS],
Eur. Phys. J. C \textbf{77}, no.12, 898 (2017)
[arXiv:1708.03247 [hep-ex]].


\bibitem{ATLAS:2017bfj}
M.~Aaboud \textit{et al.} [ATLAS],
JHEP \textbf{01}, 126 (2018)
[arXiv:1711.03301 [hep-ex]].

\bibitem{ATLAS:2019zrq}
M.~Aaboud \textit{et al.} [ATLAS],
Eur. Phys. J. C \textbf{80}, no.8, 754 (2020)
[arXiv:1903.07570 [hep-ex]].

\bibitem{ATLAS:2021kxv}
G.~Aad \textit{et al.} [ATLAS],
Phys. Rev. D \textbf{103}, no.11, 112006 (2021)
[arXiv:2102.10874 [hep-ex]].

\bibitem{ATLAS:2020dsf}
G.~Aad \textit{et al.} [ATLAS],
Eur. Phys. J. C \textbf{80}, no.8, 737 (2020)
[arXiv:2004.14060 [hep-ex]].

\bibitem{ATLAS:2020xzu}
G.~Aad \textit{et al.} [ATLAS],
JHEP \textbf{04}, 174 (2021)
[arXiv:2012.03799 [hep-ex]].

\bibitem{ATLAS:2021hza}
G.~Aad \textit{et al.} [ATLAS],
JHEP \textbf{04}, 165 (2021)
[arXiv:2102.01444 [hep-ex]].

    
\bibitem{ATLAS:2019gdh}
G.~Aad \textit{et al.} [ATLAS],
JHEP \textbf{12}, 060 (2019)
[arXiv:1908.03122 [hep-ex]].


\bibitem{ATLAS:2021pzz}
G.~Aad \textit{et al.} [ATLAS],
Phys. Rev. D \textbf{104}, no.3, 032014 (2021)
[arXiv:2103.08189 [hep-ex]].


\bibitem{ATLAS:2019fag}
G.~Aad \textit{et al.} [ATLAS],
JHEP \textbf{06}, 046 (2020)
[arXiv:1909.08457 [hep-ex]].

\bibitem{ATLAS:2021yij}
G.~Aad \textit{et al.} [ATLAS],
JHEP \textbf{05}, 093 (2021)
[arXiv:2101.12527 [hep-ex]].


\bibitem{CMS:2022rqk}
 [CMS],
[arXiv:2207.02254 [hep-ex]].

\bibitem{CMS:2022goy}
 [CMS],
[arXiv:2208.02717 [hep-ex]].

\bibitem{Baer:2022qqr}
H.~Baer, V.~Barger, X.~Tata and K.~Zhang,
Symmetry \textbf{14}, no.10, 2061 (2022)
[arXiv:2209.00063 [hep-ph]].

\bibitem{Baer:2022smj}
H.~Baer, V.~Barger, X.~Tata and K.~Zhang,
[arXiv:2212.09198 [hep-ph]].

\bibitem{ATLAS:2021ilc}
 [ATLAS],
ATL-PHYS-PUB-2021-019.	


\bibitem{ATLAS:2022ihe}
G.~Aad \textit{et al.} [ATLAS],
Eur. Phys. J. C \textbf{83}, no.7, 561 (2023)
doi:10.1140/epjc/s10052-023-11543-6
[arXiv:2211.08028 [hep-ex]].

\bibitem{Aoyama:2020ynm}
T.~Aoyama, N.~Asmussen, M.~Benayoun, J.~Bijnens, T.~Blum, M.~Bruno, I.~Caprini, C.~M.~Carloni Calame, M.~C\`e and G.~Colangelo, \textit{et al.}
Phys. Rept. \textbf{887}, 1-166 (2020)
doi:10.1016/j.physrep.2020.07.006
[arXiv:2006.04822 [hep-ph]].
\end{thebibliography}
\end{document}